\documentclass[12pt,english]{article}

\renewcommand{\vec}[1]{\mathbf{#1}}
\newcommand{\CO}{{\cal O}}
\newcommand{\CC}{{\cal C}}
\newcommand{\CF}{{\cal F}}
\newcommand{\mY}{{\mathbb Y}}

\newcommand{\so}{{\mathfrak{so}}}
\newcommand{\SO}{{\text{SO}}}

\makeatletter
\newcommand*{\rom}[1]{\expandafter\@slowromancap\romannumeral #1@}
\makeatother

\usepackage[latin9]{inputenc}
\usepackage{geometry}
\usepackage{verbatim}
\usepackage{prettyref}
\usepackage{amsmath}

\usepackage{mathtools}

\DeclarePairedDelimiter\floor{\lfloor}{\rfloor}
\usepackage{amssymb}
\usepackage{cases}
\usepackage{ytableau}
\usepackage{graphicx}
\usepackage{caption}
\usepackage{subcaption}
\usepackage{tikz}
\usepackage[vcentermath]{youngtab}

 \usepackage{slashed}
\usepackage{esvect}
\usepackage{dsfont}
\usepackage{empheq}

\usepackage{color}
\definecolor{darkgreen}{rgb}{0,0.5,0}
\definecolor{darkblue}{rgb}{0,0,0.6}
\definecolor{purple}{rgb}{0.4,.2,0.7}
\usepackage[colorlinks=true,citecolor=darkgreen,linkcolor=purple,urlcolor=purple]{hyperref}

\numberwithin{equation}{section}
\numberwithin{figure}{section}
\numberwithin{table}{section}

\def\CD{{\cal D}}
\def\tr{\,{\rm tr}\,}

\DeclareFontShape{OT1}{cmr}{mx}{n}{<->cmr10}{}

\begin{document}

\fontseries{mx}\selectfont

\begin{center}
\LARGE \bf Higher spin de Sitter quasinormal modes 
\end{center}

\vskip1cm

\begin{center}
Zimo Sun
\vskip5mm
{
\it{\footnotesize  Department of Physics, Columbia University}\\
}
\end{center}

\vskip2cm

\vskip0mm

\begin{abstract}
We construct higher spin quasinormal modes algebraically in $D$-dimensional de Sitter spacetime using the ambient space formalism. The quasinormal modes fall into two nonunitary lowest-weight representations of $\so(1, D)$. From a local QFT point of view, the lowest-weight  quasinormal modes of massless higher spin fields  are produced by gauge-invariant boundary conserved currents and boundary higher-spin Weyl tensors inserted at the  southern pole of the past boundary. We also show that the quasinormal spectrum of a massless/massive spin-$s$ field is precisely encoded in the Harish-Chandra character corresponding to the unitary massless/massive spin-$s$ $\SO(1, D)$ representation.
\end{abstract}

\newpage

\tableofcontents

\section{Introduction}
In this paper, we present an algebraic approach to the construction of higher spin quasinormal modes in de Sitter spacetime (dS). In the framework of general relativity (GR), quasinormal modes can be  defined as the damped modes of some perturbation in a classical gravitational background  with a horizon, like black holes and dS.  Astrophysically, they are important because the least damped gravitational quasinormal mode of a Schwarzschild black hole is detected and measured by LIGO through  gravitational waves emitted during the  so-called ``ringdown'' phase  \cite{TheLIGOScientific:2016src}. The measured value of gravitational quasinormal frequency can be used to test GR which predicts that the spin and mass of a black hole completely fix gravitational quasinormal frequencies.


One standard method of finding quasinormal modes in a generic  background with a horizon is solving the equation of motion for a perturbation, in most cases numerically, and then imposing  in-falling boundary condition at the horizon \cite{Natario:2004jd,Berti:2009kk,Kokkotas:1999bd,Konoplya:2011qq}. In the static patch of de Sitter spacetime, c.f. (\ref{statmetric}),  by separation of variables the radial parts of quasinormal modes are found to satisfy hypergeometric functions and hence can be solved analytically (see 
\cite{Brady:1999wd, LopezOrtega:2012vi, LopezOrtega:2006my} for a summary and derivation of the analytical results associated to 
scalars, Dirac spinors, Maxwell fields and linearized gravity in any dimensions).  The underlying reason for the existence of these analytical solutions is the large dS isometry group which organizes quasinormal modes according to certain representation structure. Such a representation structure was first discovered in \cite{Ng:2012xp,Jafferis:2013qia} for a massive scalar field with mass $m^2=\Delta(3-\Delta), 0<\Delta<3$ \footnote{The representation carried by such a scalar field is in the (scalar) complementary series.} in $\text{dS}_4$. In this case, the quasinormal modes of the scalar field comprise two (non-unitary) lowest-weight representations of the isometry algebra $\so(1, 4)$, which is also the conformal algebra of $\mathbb{R}^3$. More explicitly, the authors built two lowest-weight/primary quasinormal modes of quasinormal frequency $i\omega=\Delta$ and $i\omega=\bar\Delta=3-\Delta$ respectively, as the two leading asymptotic behaviors  of vacuum-to-vacuum bulk  two-point function when one point pushed to the northern pole on the future sphere. Upon each primary quasinormal mode, an infinite tower of quasinormal modes can be generated as $\so(1,4)$-descendants and the scaling dimension of every descendant can be identified as $(i\times\text{quasinormal frequency})$. The two towers of quasinormal modes together span the whole  scalar quasinormal spectrum. These results were later reformulated by \cite{Tanhayi:2014kba} in the ambient space formalism  and generalized to  massive vector fields and Dirac spinors in the same paper. We'll call the way of constructing quasinormal modes using the dS isometry group as in \cite{Ng:2012xp,Jafferis:2013qia,Tanhayi:2014kba} the {\it algebraic} approach.

The separation of variables method in \cite{Brady:1999wd,  LopezOrtega:2006my}   is increasingly cumbersome when applied to higher spin fields due to the rapidly increasing number of tensor structures. So in this paper we will focus on generalizing the algebraic approach to construct quasinormal modes of higher spin fields, which are formulated in the ambient space (see section \ref{amb} for a review about the ambient space formalism).
In section \ref{construction}, we first review the algebraic construction of quasinormal modes of a scalar field $\varphi$ of mass $m$ in $\text{dS}_{d+1}$ using the ambient space formalism.
The quasinormal spectrum found in this way can be packaged into a ``{\it quasinormal character}'', cf. (\ref{quacha})
\begin{align}
\chi^{\text{QN}}\equiv \sum_\omega d_\omega\,q^{i\omega}
\end{align}
where the sum runs over all quasinormal frequencies and $d_\omega$ is the degeneracy of quasinormal modes with frequency $\omega$.
We show that the quasinormal character of $\varphi$ is given by
\begin{align}\label{intros}
\chi^{\text{QN}}_\varphi(q)=\frac{q^\Delta+q^{\bar\Delta}}{(1-q)^d},\,\,\,\,\,\, \bar\Delta=d-\Delta
\end{align}
where $\Delta=\frac{d}{2}+\sqrt{\frac{d^2}{4}-m^2}$ is the scaling dimension of $\varphi$. According to \cite{Basile:2016aen,Dobrev:1977qv,Hirai:Char}, $\chi^{\text{QN}}_\varphi(q)$ is the Harish-Chandra $\SO(1, d+1)$ character corresponding to the unitary (scalar) principal series when $m>\frac{d}{2}$ and the unitary (scalar) complementary series when $0<m\le \frac{d}{2}$. The $(\Delta\leftrightarrow\bar\Delta)$ symmetry in (\ref{intros}) manifests the {\it two} towers of quasinormal modes. The generalization of the algebraic construction to  {\it massive}  higher spin fields is straightforward. The only difference is that the two primary quasinormal modes have spin degeneracy. In this case, the quasinormal character is given by eq. (\ref{quachar1}).
However, the generalization to  the {\it massless} higher spin case is quite nontrivial because gauge symmetry significantly reshapes quasinormal spectrums compared to the massive case. For example, the naive $(\Delta\leftrightarrow\bar\Delta)$ symmetry would lead to growing modes instead of damped modes because $\bar\Delta=2-s<0$ when $s\ge 3$. Moreover the symmetry disagrees with the result of \cite{LopezOrtega:2006my} for $s=1, 2$. On the representation side, the underlying reason for the difficulty in generalization is that the massless higher spin fields are in the exceptional series for $d\ge 4$ and in the discrete series for $d=3$ while generic massive fields are in the principal series or the complementary series \footnote{We exclude the partially massless fields while talking about massive fields.}. In section \ref{spinning}, we'll discuss the subtleties associated to gauge symmetry in more detail and explain how to take into account gauge symmetry properly while constructing {\it physical} quasinormal modes. In particular, the two-tower structure still holds and the two primary quasinormal modes are given by eq.(\ref{lw}) and eq.(\ref{lw2}). In addition, in section \ref{southernpole}, we argue that the two primary quasinormal modes are produced by insertions of boundary gauge-invariant conserved currents (of scaling dimension $d+s-2$) and boundary higher-spin Weyl tensors (of scaling dimension 2) at the southern pole of the past sphere (see fig. {\ref{QFTPOV}). Other quasinormal modes are sourced by the descendants of these two operators.

\begin{figure}[h]
\centering
  \includegraphics[width=0.4\linewidth]{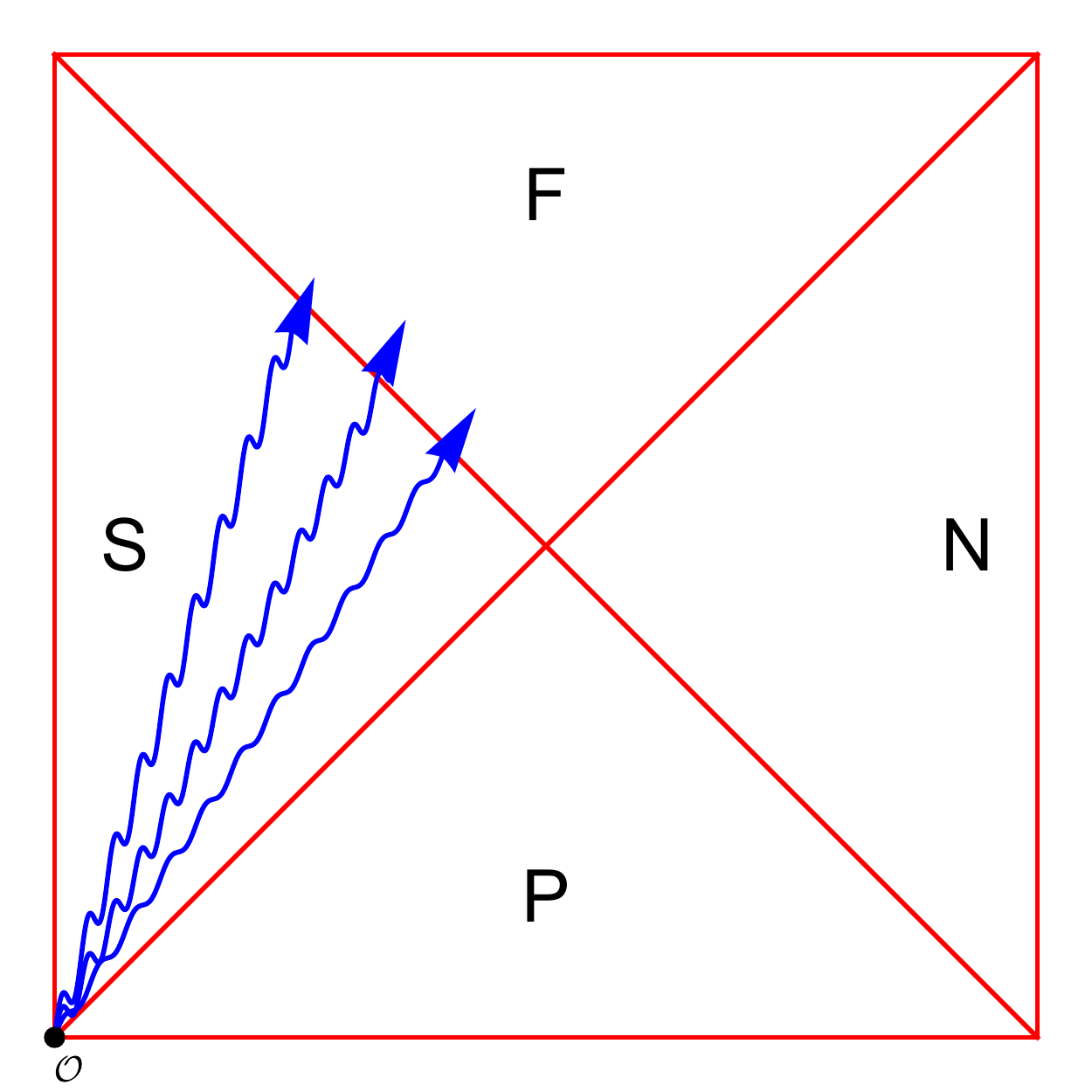}
  \caption{\small The Penrose diagram of de Sitter spacetime. Quasinormal modes (in the southern static patch ``S'') of massless higher spin fields are sourced by certain gauge-invariant operators $\CO$  inserted at the southern pole of the past sphere.}
  \label{QFTPOV}
\end{figure}

Based on the algebraic construction described in section \ref{spinning}, we extract the physical quasinormal spectrum of  massless higher spin fields in section \ref{char} and compute the corresponding quasinormal character.  Here, we list some examples at $d=3,4 ,5$ (see eq. (\ref{flipping}) for a  general expression working in any $d$):
\small
\begin{align}
&d=3:\,\,\,\,\,\,\,\,\,\,\,\,\,\,\,\,\,\,\,\,\,\,\,\,\,\,\,\,\,\,\,\,\,\,\,\,\,\,\,\,\,\,\,\,\,\,\,\, \chi^{\text{QN}}_s(q)=2\, \frac{(2s+1)\,q^{1+s}-(2s-1)\,q^{2+s}}{(1-q)^3}\nonumber\\
&d=4:\,\,\,\,\,\,\,\,\,\,\,\,\,\,\,\,\,\,\,\,\,\,\,\,\,\,\,\,\,\,\,\,\,\,\,\,\,\,\,\,\,\,\,\,\,\,\,\,\,\,\,\,\,\,\,\,\,\,\,\,\,\,\,\,\,\,\,\, \chi^{\text{QN}}_s(q)=2\, \frac{(2s+1)\, q^2}{(1-q)^4}\\
&d=5:\,\,\,\, \chi^{\text{QN}}_s(q)=\frac{1}{3}(s+1)(2s+1)(2s+3)\frac{q^2-q^3}{(1-q)^5}+\frac{s+1}{3}\frac{(s+2)(2s+3)q^{s+3}-s(2s+1)q^{s+4}}{(1-q)^5}\nonumber
\end{align}
\normalsize
These quasinormal characters coincide with the original computation of  Harish-Chandra $\SO(1,d+1)$ characters in  \cite{Hirai:Irrep,Hirai:Char} and the characters appearing in the one-loop path integral of massless higher spin fields on $S^{d+1}$, the Wick rotation of $\text{dS}_{d+1}$ \cite{Anninos:2020hfj}. Therefore, the quasinormal characters, which are defined in a pure physics setup, connect {\it nonunitary} lowest-weight representations of $\so(1,d+1)$ and the {\it unitary} representations of $\SO(1,d+1)$. 
In addition, in the appendix \ref{match}, we recover the quasinormal spectrum of Maxwell theory and linearized gravity in \cite{LopezOrtega:2006my} by using $\chi^{\text{QN}}_1(q)$ and $\chi^{\text{QN}}_2(q)$.

When $d\ge 4$, the expansion of quasinormal character $\chi^{\text{QN}}_s(q)$ always starts from a $q^2$ term because it corresponds to a boundary higher-spin Weyl tensor insertion.  
When  $d=3$, $\chi^{\text{QN}}_s(q)$ starts from $q^{1+s}$ since the spin-$s$ Weyl tensor, which vanishes identically on the 3-dimensional boundary, gets replaced by the co-called Cotton tensor \cite{Cotton:original,Cotton:1,TD,Henneaux:2015cda}, which involves $(2s+1)$ derivatives on the boundary gauge field of scaling dimension $2-s$.


\section{Ambient space formalism for fields in de Sitter}\label{amb}
In this section, we review ambient space formalism for higher spin fields in $(d+1)$-dimensional de Sitter spacetime, which is realized as a hypersurface 
\begin{align}\label{hyper}
\eta_{AB}X^A X^B=1, \,\,\,\,\,  \eta_{AB}=(-,+,\cdots,+)
\end{align}
in the ambient space $\mathbb{R}^{1,d+1}$, where $A=0,1,\cdots, d+1$. Local intrinsic coordinates $y^\mu$ are defined through  an embedding $X^A(y)$ that satisfies (\ref{hyper}) and such an embedding induces a local metric $ds^2=g_{\mu\nu}dy^\mu dy^\nu$ on $\text{dS}_{d+1}$. To gain some intuitions about the ambient space description in field theory,  let's consider a free  scalar field $\varphi(y)$, defined in the local coordinates $y^\mu$, of mass $m^2=\Delta(d-\Delta)$  and satisfying equation of motion: $\nabla^2\varphi=m^2\varphi$ with $\nabla^2$ being the scalar Laplacian. This scalar field $\varphi$ admits a unique extension $\phi(X)$ to the ambient space such that
\begin{align}\label{extcon}
\phi(\lambda X)=\lambda^{-\kappa} \phi(X), \,\,\,\,\, \phi(X(y))=\varphi(y)
\end{align}
where $\kappa$ is an arbitrary constant. Define radial coordinate $R=\sqrt{X^2}$ and hence any point in the ambient space can be parameterized by $(R, y^\mu)$ via a dS foliation. In terms of the radial coordinate, the extension condition (\ref{extcon}) can be rephrased as $\phi(X)=\phi(R, y^\mu)=R^{-\kappa}\varphi(y)$ and the ambient space Laplacian can be expressed as $\partial_X^2=\frac{1}{R^{d+1}}\partial_R (R^{d+1}\partial_R)+\frac{1}{R^2}\nabla^2$, which together with the equation of motion of $\varphi$, yields 
\begin{align}
\partial_X^2\phi=\frac{1}{R^2}(\Delta(d-\Delta)-\kappa(d-\kappa))\phi
\end{align}
In particular, if we choose $\kappa$ to be  $\Delta$ or $\bar\Delta$, $\phi(X)$ becomes a harmonic function in the ambient space. Altogether, the scalar field $\varphi(y)$ in $\text{dS}_{d+1}$ of scaling dimension $\Delta$ is equivalent to a homogeneous harmonic  function $\phi(X)$ in the ambient space $\mathbb{R}^{1,d+1}$, i.e. 
\begin{align}\label{scalaramb}
(X\cdot\partial_X+\kappa)\phi(X)=\phi(X), \,\,\,\,\,\partial_X^2\phi(X)=0
\end{align}
where $\kappa=\Delta$ or $\kappa=\bar\Delta$. The obvious technical advantage of this ambient space description is replacing the cumbersome covariant derivative $\nabla_\mu$ by the simple ordinary derivative $\partial_{X^A}$. Such a simplification is more crucial when we deal with higher spin fields. One can find a very good review about the ambient space formalism for spinning fields in AdS in 
\cite{Sleight:2017krf,Sleight:2017cax}. Here we present an adapted version of the ambient space description in dS.

\subsection{Higher spin fields in ambient space formalism}
The totally symmetric transverse (on-shell) spin-$s$ field $\varphi_{\mu_1...\mu_s}(y)$ of scaling dimension $\Delta$ in $\text{dS}_{d+1}$ is represented uniquely in ambient space by the symmetric tensor $\phi_{A_1...A_s}(X)$,
\begin{align}\label{pullback}
\varphi_{\mu_1\cdots\mu_s}(y)=\frac{\partial X^{A_1}}{\partial y^{\mu_1}}\cdots \frac{\partial X^{A_s}}{\partial y^{\mu_s}}\phi_{A_1\cdots A_s}(X)
\end{align}
 satisfying the following equations:
\begin{itemize}
\item Tangentiality to surfaces of constant $R=\sqrt{X^2}$:
\begin{align}\label{tan}
(X\cdot \partial_U)\, \phi_s(X, U)=0
\end{align}
\item The homogeneity condition:
\begin{align}\label{homo}
(X\cdot \partial_X +\kappa)\phi_s(X, U)=0
\end{align}
A convenient choice is $\kappa=\Delta$ or $\kappa=\bar\Delta$ because, as we've seen in the scalar case, it yields the simplest equation of motion as follows:
 \item The Casimir condition, i.e. equation of motion 
\begin{align}\label{cas}
(\partial_X\cdot\partial_X)\, \phi_s(X, U)=0
\end{align}
\item The transverse condition:
\begin{align}\label{trans}
(\partial_X\cdot\partial_U)\, \phi_s(X, U)=0
\end{align}
\item The traceless condition:
\begin{align}\label{trace}
(\partial_U\cdot\partial_U)\, \phi_s(X, U)=0
\end{align}
\end{itemize}
where we've used the generating function $\phi_s(X, U)\equiv \frac{1}{s!} \phi_{A_1\cdots A_s}(X)U^{A_1}\cdots U^{A_s}$ with $U^A$ being a constant auxiliary vector. The first two conditions ensure that $\phi_s(X, U)$ is the unique uplift of $\varphi_{\mu_1\cdots\mu_s}$ that satisfies (\ref{pullback}) and the last three conditions are equivalent to the Fierz-Pauli system:
\begin{align}
&\text{The Casimir condition}: \,\,\,\,\,\,\,\,\,\, \nabla^2\varphi_{\mu_1\cdots\mu_s}=(\Delta(d-\Delta)+s)\varphi_{\mu_1\cdots\mu_s}\\
&\text{The transverse condition}: \,\,\,\,\nabla^{\mu_1}\varphi_{\mu_1\cdots\mu_s}=0\\
&\text{The traceless condition}: \,\,\,\,\,\,\,\,\, g^{\mu_1\mu_2}\varphi_{\mu_1\cdots\mu_s}=0
\end{align}
In the remaining part of the paper, we'll call (\ref{tan})-(\ref{homo}) the {\it uplift conditions} and (\ref{cas})-(\ref{trace}) the {\it Fierz-Pauli conditions}.


When $\varphi_{\mu_1\cdots\mu_s}$ is a massless spin-$s$ bulk field, the uplift conditions and Fierz-Pauli conditions have a gauge symmetry, with the gauge transformation takes the following simple form if we choose $\boxed{\kappa=2-s=\bar\Delta}$ in eq. (\ref{homo}):
\begin{align}\label{gt}
\delta_{\xi_{s-1}}\phi_s(X, U)=(U\cdot \partial_X)\,\xi_{s-1}(X, U)
\end{align}
where $\xi_{s-1}$ satisfies 
\begin{itemize}
\item Tangentiality to surfaces of constant $R=\sqrt{X^2}$:
\begin{align}\label{tanx}
(X\cdot \partial_U)\, \xi_{s-1}(X, U)=0
\end{align}
\item The homogeneity condition:
\begin{align}\label{homox}
 (X\cdot\partial_X+1-s)\xi_{s-1}(X, U)=0
\end{align}
\item The Casimir condition:
\begin{align}\label{casx}
(\partial_X\cdot\partial_X)\, \xi_{s-1}(X, U)=0
\end{align}
\item The transverse condition:
\begin{align}\label{transx}
(\partial_X\cdot\partial_U)\, \xi_{s-1}(X, U)=0
\end{align}
\item The traceless condition:
\begin{align}\label{tracex}
(\partial_U\cdot\partial_U)\, \xi_{s-1}(X, U)=0
\end{align}
\end{itemize}
In the intrinsic coordinate language, this set of  equations implies that $\xi_{s-1}$ is a transverse traceless symmetric spin-$(s-1)$ field on $\text{dS}_{d+1}$ satisfying on-shell equation of motion $(\nabla^2-(s-1)(s+d-2))\xi_{s-1}=0$, where spin indices are suppressed.

\subsection{Isometry group in ambient space formalism}
$\text{dS}_{d+1}$ is a maximally symmetric space with isometry group $\SO(1,d+1)$ generated by $L_{AB}=-L_{BA}$, subject to the commutation relations:
\begin{align}
[L_{AB}, L_{CD}]=\eta_{BC}L_{AD}+\eta_{DB}L_{CA}+\eta_{AD}L_{BC}+\eta_{CA}L_{DB}
\end{align}
For unitary representations,  $L_{AB}$ are anti-hermitian. The following linear combinations of $L_{AB}$
\begin{align}\label{DtoL}
D=L_{0, d+1},\,\,\, M_{ij}=L_{ij},\,\,\, P_i=L_{d+1,i}+L_{0,i},\,\,\, K_i=L_{d+1,i}-L_{0,i}
\end{align}
lead to the  conformal algebra of $\mathbb{R}^d$. Here we list some nontrivial commutators of the conformal algebra
\begin{align}\label{commu}
&[D, P_i]=P_i,\,\,\,\, [D, K_i]=-K_i,\,\,\,\, [K_i, P_j]=2\delta_{ij} D- 2M_{ij}\nonumber\\
&[M_{ij}, P_{k}]=\delta_{jk}P_i-\delta_{ik}P_j, \,\,\,\, [M_{ij}, K_{k}]=\delta_{jk}K_i-\delta_{ik}K_j
\end{align}

$\SO(1,d+1)$ acts linearly on  fields in the ambient space $\mathbb{R}^{1,d+1}$. In particular, the generators $L_{AB}$ are realized as linear differential operators in both $X$ and $U$:
\begin{align}\label{LAD}
L_{AB}=(X_A\partial_{X^B}-X_B\partial_{X^A})+(U_A\partial_{U^B}-U_B\partial_{U^A})
\end{align}
where the first term corresponds to orbital angular momentum and the second term represents spin angular momentum.
The action of $M_{ij}, P_i, K_i, D$ induced by (\ref{LAD}) is
\begin{align}\label{isoalg}
&M_{ij}=X_i\partial_{X^j}-X_j\partial_{X^i}+U_i\partial_{U^j}-U_j\partial_{U^i}\\
&K_i=X^+ \partial_{X^i}+2X_i\partial_{X^-}+U^+ \partial_{U^i}+2U_i\partial_{U^-}\\
&P_i=-X^-\partial_{X^i}-2X_i\partial_{X^+}-U^-\partial_{U^i}-2U_i\partial_{U^+}\\
&D=-X^+\partial_{X^+}+X^-\partial_{X^-}-U^+\partial_{U^+}+U^-\partial_{U^-}
\end{align}
where we've used lightcone coordinates $X^\pm\equiv X^0\pm X^{d+1}$ and $U^\pm\equiv U^0\pm U^{d+1}$.

 With a little algebra, one can show that all $L_{AB}$ commute with the following set of differential operators:
  \begin{align}
  X\cdot\partial_U, \,\,\,\,\,X\cdot\partial_X, \,\,\,\,\,\partial_X^2,\,\,\,\,\,\partial_X\cdot\partial_U,\,\,\,\,\, \partial_U^2, \,\,\,\,\,U\cdot\partial_X
  \end{align}
 The first five operators define the uplift conditions and Fierz-Pauli conditions, cf. (\ref{tan})-(\ref{trace}), and hence the commutation relations imply that the on-shell bulk fields carry representations of $\so(1,d+1)$. The last operator defines the gauge transformation of a massless higher spin field and therefore one implication of $[L_{AB},U\cdot\partial_X]=0$ is that the descendants of a pure gauge mode are also pure gauge. This observation is crucial when we construct quasinormal modes for massless higher spin fields.

\section{Algebraic construction of quasinormal modes}\label{construction}
The southern static patch of de Sitter spacetime, which corresponds to the region denoted by ``S'' in the fig. (\ref{QFTPOV}), has coordinates 
\begin{align}\label{statcoord}
X^0=\sqrt{1-r^2}\sinh t, \,\,\,\,\, X^i=r \,\Omega^i,\,\,\,\,\, X^{d+1}= \sqrt{1-r^2}\cosh t
\end{align}
and shows a manifest spherical horizon at $r=1$ or $\rho=\infty$ in its metric
\begin{align}\label{statmetric}
ds^2&=-(1-r^2)dt^2+\frac{dr^2}{1-r^2}+r^2 d\Omega^2\nonumber\\
&=\frac{-dt^2+d\rho^2}{\cosh^2\rho}+\tanh^2\rho\, d\Omega^2
\end{align}
where $r=\tanh\rho $ and $d\Omega^2=h_{ab}d\vartheta^a d\vartheta^b$ is the standard metric on $S^{d-1}$ ($\vartheta^a$ are spherical coordinates on $S^{d-1}$). The traditional analytical approach to quasinormal requires solving the equation of motion in bulk and imposing  in-falling boundary condition, i.e. $e^{-i\omega_{\text{QN}}(t-\rho)}$, for the leading asymptotic behavior near the horizon at $\rho=\infty$. 

An algebraic method of solving quasinormal modes  was first used for scalar fields in $\text{dS}_4$ in \cite{Ng:2012xp, Jafferis:2013qia}. In particular, the authors found that all the quasinormal modes fall into two lowest-weight representations of the conformal algebra $\so(1,d+1)$. Therefore, it suffices to find the two lowest-weight/primary quasinormal modes, which are solutions to the equation of motion and are annihilated by $K_i$, and the rest of the quasinormal spectrum can be generated as descendants of them. In this section, we will first reformulate this scalar story using the ambient space formalism  and then generalize it to  higher spin fields.

\subsection{Scalar fields}
Let $\varphi(X)$ be a free scalar field of mass $m^2=\Delta(d-\Delta)>0$ in $\text{dS}_{d+1}$. By construction, the equation of motion $(\nabla^2-m^2)\varphi=0$ is satisfied by the boundary-to-bulk propagators, which in ambient space take the following form:
\begin{align}\label{anb}
\alpha_\Delta(X;\xi)=\frac{1}{(X\cdot\xi)^\Delta},\,\, \,\,\, \beta_{\Delta}(X;\xi)=\frac{1}{(X\cdot\xi)^{\bar\Delta}}
\end{align}
where $\xi^A$ is a constant null vector in $\mathbb{R}^{1,d+1}$ representing a point on the future/past boundary of $\text{dS}_{d+1}$.  Treating $\alpha_\Delta$ and $\beta_\Delta$ as mode functions in $X^A$, they are primary with respect to the conformal algebra $\so(1,d+1)$ if $X\cdot\xi=X^+$, because $K_i$ only involves derivatives $\partial_{X^i}$ and $\partial_{X^-}$ while acting on scalar fields (cf. (\ref{isoalg})). By choosing $\xi^A=(-1,0,\cdots,0, 1)$ which is the southern pole of the past sphere, we obtain the two primary quasinormal modes
\begin{align}
&\alpha_\Delta(X)=\frac{1}{(X^+)^\Delta}= (\cosh\rho)^\Delta e^{-\Delta \, t}\overset{\rho\to \infty}{\rightsquigarrow} e^{-\Delta(t-\rho)}\nonumber\\
&\beta_\Delta(X)=\frac{1}{(X^+)^{\bar\Delta}}= (\cosh\rho)^{\bar\Delta} e^{-\bar\Delta \, t}\overset{\rho\to \infty}{\rightsquigarrow} e^{-\bar\Delta(t-\rho)}
\end{align}
with quasinormal frequency $i\omega_\alpha=\Delta$ and $i\omega_\beta=\bar\Delta$ respectively. The rest quasinormal modes can be realized as descendants of $\alpha_\Delta(X)$ and $\beta_\Delta(X)$. Though not explicitly spoken out in \cite{Ng:2012xp, Jafferis:2013qia}, this claim actually relies on two facts: (a) $P_i$ preserves the equation of motion and (b) $P_i$ preserves the in-falling boundary condition near horizon. The former is obvious as we've seen at the end of last section that the $\SO(1,d+1)$ action preserves uplift conditions and Pauli-Fierz conditions. The latter holds because $P_i$ is dominated by $-2 X_i \partial_{X^+}$ near horizon, where $X_i\approx \Omega_i$ and $X^+\approx 2 \, e^{t-\rho}$. With the quasinormal modes known, we need to figure out the corresponding frequency. This is quite straightforward in our formalism. By construction, each quasinormal mode is an eigenfunction of the dilatation operator $D$, which is just $-\partial_t$ in the static patch. Using the in-falling boundary condition $e^{-i\omega_{\text{QN}}(t-\rho)}$ near horizon, we can identify the scaling dimension, i.e. eigenvalue with respect to $D$, as $i\times(\text{quasinormal frequency}\,\, \omega_{\text{QN}})$. For example, a descendant of $\alpha_\Delta$  at level $n$ is a quasinormal mode of frequency $\omega_{\text{QN}}=- i(\Delta+n)$ and similarly for $\beta_\Delta$.

To end the discussion about scalar quasinormal modes, let's compare our construction with the known result in literature.  For example, in \cite{LopezOrtega:2006my}, the scalar quasinormal modes in $\text{dS}_{d+1}$ are found to be
\begin{align}
\varphi^{\text{QN}}_{\omega}(t, r,\Omega)=r^\ell (1-r^2)^{\frac{i\omega}{2}} F\left(\frac{\ell+i\omega+\Delta}{2},\frac{\ell+i\omega+\bar\Delta}{2}, \frac{d}{2}+\ell, r^2\right) Y^{\ell\sigma}(\Omega) e^{-i\omega t}
\end{align}
where $Y^{\ell\sigma}(\Omega)$ denote spherical harmonics on $S^{d-1}$ and the quasinormal frequency $\omega$ takes the following values
\begin{align}\label{specscalar}
\omega_{\ell, n}=-i(\Delta+\ell+2n),\,\,\,\,\, \bar\omega_{\ell, n}=-i(\bar\Delta+\ell+2n),\,\,\,\,\, \ell, n\in\mathbb N
\end{align}
For  fixed $\ell$ and $n$, the quasinormal modes of frequency $\omega_{\ell,n}$ or $\bar\omega_{\ell,n}$ have degeneracy $D^d_\ell$, i.e. the dimension of the spin-$\ell$ representation of $\SO(d)$. 
In particular, the two quasinormal modes corresponding to $\ell=n=0$ are
\begin{align}
&\varphi^{\text{QN}}_{\omega_{0,0}}=\frac{1}{\sqrt{A_{d-1}}}\frac{e^{-\Delta t}}{(1-r^2)^{\frac{\Delta}{2}}}=\frac{1}{\sqrt{A_{d-1}}}(\cosh\rho)^\Delta e^{-\Delta t}\nonumber\\
&\varphi^{\text{QN}}_{\bar\omega_{0,0}}=\frac{1}{\sqrt{A_{d-1}}}\frac{e^{-\bar\Delta t}}{(1-r^2)^{\frac{\bar\Delta}{2}}}=\frac{1}{\sqrt{A_{d-1}}}(\cosh\rho)^{\bar\Delta} e^{-\bar\Delta t}
\end{align}
where $A_{d-1}$ is the area of $S^{d-1}$. Apart from the normalization constant, these two quasinormal modes are exactly the primary quasonormal modes $\alpha_\Delta$ and $\beta_\Delta$ respectively. In addition to the match of the primary quasinormal modes, we can also show that the algebraic construction reproduces the quasinormal spectrum (\ref{specscalar}). Define $P=\sqrt{P_i P_i}$ and $\hat P_i =P^{-1} P_i$. Then the linear independent descendants of $\alpha_\Delta$ are of the form $P^{2n+\ell} Y^{\ell\sigma}(\hat P)\alpha_\Delta$ with $\ell,n\in\mathbb N$. For fixed $\ell$ and $n$, these are quasinormal modes corresponding to $\omega_{\ell,n}$. Similarly $P^{2n+\ell} Y^{\ell\sigma}(\hat P)\beta_\Delta$ represents quasinormal modes corresponding to $\bar\omega_{\ell,n}$. 

\subsection{Massive higher spin fields}\label{spinning1}
As in the scalar case, we need to start from a solution of the Pauli-Fierz conditions (\ref{cas})-(\ref{trace}), subject to the tangential condition and homogeneous condition. A natural candidate is the higher spin boundary-to-bulk propagator. In AdS, the boundary-to-bulk propagator of a spin-$s$ field with a generic scaling dimension $\Delta(\not=d+s-2)$ is given by \cite{Mikhailov:2002bp,Costa:2014kfa, Sleight:2017cax}
\begin{align}\label{bbK}
K^{\text{AdS}}_{[\Delta, s]}(X, U; \xi, Z)=\frac{\left[(U\cdot Z)(\xi\cdot X)-(U\cdot \xi)(Z\cdot X)\right]^s}{(X\cdot \xi)^{\Delta+s}}
\end{align}
where the null vector $\xi\in \mathbb{R}^{1, d+1}$ represents a boundary point and the null vector $Z\in\mathbb C^{1, d+1}$, satisfying $\xi\cdot Z=0$, encodes the {\it boundary} spin.  $K^{\text{AdS}}_{[\Delta, s]}$ in (\ref{bbK}) scales like $K^{\text{AdS}}_{[\Delta, s]}(\lambda X)=\lambda^{-\Delta}K^{\text{AdS}}_{[\Delta, s]}(X)$, which is the analogue of  $\kappa=\Delta$ in eq. (\ref{homo}). In AdS, this scaling property corresponds to the choice of ordinary boundary condition. In dS, on the other hand, both near-boundary fall-offs of a bulk field are dynamical and hence there are two boundary-to-bulk propagators 
\begin{align}\label{bbK}
K^{\text{dS}}_{[\kappa, s]}(X, U; \xi, Z)=\frac{\left[(U\cdot Z)(\xi\cdot X)-(U\cdot \xi)(Z\cdot X)\right]^s}{(X\cdot \xi)^{\kappa+s}}
\end{align}
where $\kappa\in\{\Delta,\bar\Delta\}$. Notice that $K_i$ in (\ref{isoalg}) doesn't involve any derivative with respect to $X^+$ or $U^+$. So we can obtain primary  mode functions from $K^{\text{dS}}_{[\kappa, s]}(X, U; \xi, Z)$ by putting $\xi^A$ at the southern pole of the past sphere, i.e. $\xi^A=(-1,0,\cdots,0,1)$ and choosing $Z^A=(0, z_i,0)$, where $z_i$ itself is a null vector in $\mathbb{C}^d$:
\begin{align}
\alpha_{[\Delta, s]}=\frac{\Phi^s}{(X^+)^\Delta}, \,\,\,\,\, \beta_{[\Delta, s]}=\frac{\Phi^s}{(X^+)^{\bar\Delta}}
\end{align}
where $\boxed{\Phi=X^+ u\cdot z-U^+ x\cdot z}$ (despite the lower case $x$ and $u$, indeed $x^i\equiv X^i$ and $u^i\equiv U^i$). 
$\alpha_{[\Delta, s]}$ and $\beta_{[\Delta, s]}$ are clearly primary quasinormal modes since  in-falling boundary condition near horizon naturally follows from the lack of dependence on $X^-$ and $U^-$. Given the two primary quasinormal modes, the whole quasinormal spectrum can be generated by acting $P_i$ on them repeatedly. In this sense, the algebraic construction of quasinormal modes for massive higher spin fields is a straightforward generalization of the scalar case. Before moving to  massless higher spin fields, we want to emphasize that rigorously speaking, ``$\alpha_{[\Delta, s]}$'' or ``$\beta_{[\Delta, s]}$'' is {\it not one}  primary quasinormal mode because varying $z^i$ would yield different quasinormal modes. Indeed, $\alpha_{[\Delta, s]}$ represents a collection of quasinormal modes with the same frequency  and the vector space spanned by these quasinormal modes furnishes a spin-$s$ representation of $\SO(d)$. But for convenience, in most part of the paper, we'll stick to the misnomer by calling, say $\alpha_{[\Delta, s]}$, {\it the} $\alpha$-mode.

\subsection{Massless higher spin fields}\label{spinning}
 When $\Delta$ hits $d-2+s$, i.e. the massless limit, $K^{\text{AdS}}_{[\Delta, s]}$ still holds as a boundary-to-bulk propagator in the so-called de Donder gauge  \cite{Mikhailov:2002bp, Sleight:2017cax}. So we can extract de Sitter primary quasinormal modes, in de Donder gauge, from $K^{\text{AdS}}_{[\Delta, s]}$ as in the massive case:
 \begin{align}\label{alphabeta}
&\alpha\text{-mode}: \alpha^{(s)}(X,U;z) =\frac{\Phi^s }{(X^+)^{2}}\, \left(\frac{R}{X^+}\right)^{d+2(s-2)}\nonumber\\
&\beta\text{-mode}: \beta^{(s)}(X,U;z) =\frac{\Phi^s}{(X^+)^2}
\end{align}
where the $\SO(1,d+1)$-invariant $R=\sqrt{X^2}$ is inserted in $\alpha^{(s)}$ \footnote{Including $R^{d+2(s-2)}$ doesn't spoil the Fierz-Pauli conditions (\ref{cas})-(\ref{trace}).} so that it can have the same scaling property as $\beta^{(s)}$, which corresponds to $\kappa=2-s$ in (\ref{homo}). With this choice of $\kappa$, gauge transformation acts in the same way on both modes, schematically $\delta\alpha^{(s)}=U\cdot\partial_X(\cdots)$ and $\delta\beta^{(s)}=U\cdot\partial_X(\cdots)$.

 Naively, one would expect (\ref{alphabeta}) to be the end of story since we can generate the rest quasinormal modes as descendants of $\alpha^{(s)}$ and $\beta^{(s)}$, just as in the massive case. However, this expectation is only partially correct because, as we'll show in the following, the quasinormal spectrum is significantly affected by gauge symmetry in the massless case compared to its massive counterpart. For example, the $\beta^{(s)}$-mode has quasinormal frequency $i\omega=2-s$, which would lead to an exponentially growing rather than damped behavior at future for $s\ge 3$. Gauge symmetry should be the only cure for this pathological growth and indeed, we do find that the $\beta$-mode is pure gauge for any $s\ge 1$:
\begin{align}\label{defxi}
\beta^{(s)}=U\cdot\partial_X\left(\xi_{s-1}\right), \,\,\,\,\, \xi_{s-1}=\Phi^{s-1} \frac{x\cdot z}{X^+}
\end{align}
where the gauge parameter $\xi_{s-1}$ can be realized as a descendant of another mode in the following sense:
\begin{align}\label{defeta}
\xi_{s-1}=(z\cdot P)\eta_{s-1}, \,\,\,\,\,\eta_{s-1}=-\frac{1}{2}\Phi^{s-1}\,\log X^+
\end{align}
As a result, all the quasinormal modes in the $\beta$-tower are {\it unphyical}. The $\alpha$-mode itself is not pure gauge but it has some pure-gauge descendants:
\begin{align}\label{PCD}
P\cdot\CD\,  \alpha^{(s)} &=s(s+d-3)\,   U\cdot\partial_X\left[ \Phi^{s-1}\left(\frac{R}{X^+}\right)^{d+2s-2}\right]
\end{align}
where $\CD_i=(\frac{d-2}{2}+z\cdot\partial_z) \partial_{z_i}-\frac{z_i}{2}\partial_z^2$ \cite{Dobrev:1975ru} strips off $z_i$ while respecting its nullness. If we write out the indices explicitly, (\ref{PCD}) means  $P_{i_1}\alpha^{(s)}_{i_1\cdots i_s}(X, U)=0$ up to gauge transformation, which is the reminiscence of a spin-$s$ conserved current.

In the remaining part of this section, we'll show that although the $\beta$-tower of quasinormal modes gets killed by gauge transformation, there exist a brand new tower of physical quasinormal modes.

\subsubsection{Maxwell fields}
The primary $\beta$-mode of a massless spin-1 field is $\beta^{(1)}_i=\frac{\Phi_i}{(X^+)^2}$, where $\Phi_i=X^+ u_i- U^+ x_i$. It is pure gauge and the corresponding gauge parameter takes a very special form
\begin{align}
\beta^{(1)}_i=U\cdot \partial_X \left(P_i \, \eta_0\right)
\end{align}
where $\eta_0$ is given by eq. (\ref{defeta}) with $s=1$. Since $\so(1,d+1)$ action commutes with gauge transformation, we can switch the order of $P_i$ and $U\cdot \partial_X$ in $\beta^{(1)}_i$:
\begin{align}\label{beta1}
\boxed{\beta^{(1)}_i=P_i \left(U\cdot \partial_X\, \eta_0\right)}
\end{align}
This new expression of $\beta^{(1)}_i$ inspires the following crucial observation.
Treating $\beta^{(1)}_i$ as a vector field indexed by  $i$  and treating $P_i$ as an ordinary derivative like $\partial_i$, then $\beta^{(1)}_i$ can be thought as a ``pure gauge'' mode with the gauge parameter being $U\cdot \partial_X \,\eta_0$. (We want to emphasize that this gauge symmetry structure  is completely different from the bulk gauge symmetry, which takes the form $\delta(\cdots)=U\cdot\partial_X (\cdots)$. To distinguish it from the bulk gauge symmetry, we call it a ``pseudo'' gauge symmetry and its connection with the boundary gauge transformation will be discussed in section \ref{Max}). Since $\beta^{(1)}_i$ is pure gauge with respect to the pseudo gauge symmetry, the (pseudo) field strength $\CF_{ij}\equiv P_i \beta^{(1)}_j-P_j\beta^{(1)}_i$ vanishes identically. 
The vanishing of $\CF_{ij}$ signals a potential way to obtain the new physical quasinormal modes, which will be implemented step by step as follows:
\begin{itemize}
\item First, we define a different $\beta$-mode $\hat\beta^{(1)}_i$ by deforming the scaling dimension of $\beta^{(1)}_i$ from 1 to $\Delta-1$:
\begin{align}
\hat \beta^{(1)}_i(X, U)\equiv \frac{\Phi_i}{(X^+)^\Delta}=\left(X^+\right)^{2-\Delta}\,\beta^{(1)}_i,\,\,\,\,\, \Delta\not= 2
\end{align} 
From the bulk field theory point of view, it amounts to giving a mass term to the Maxwell field to break the {\it bulk} $U(1)$ gauge symmetry. 
\item Then the new pseudo field strength $\hat \CF_{ij}\equiv P_i \hat\beta^{(1)}_j-P_j\hat\beta^{(1)}_i$ is nonvanishing
\begin{align}
\hat \CF_{ij}=2(\Delta-2)\frac{x_i\, u_j - u_i\, x_j}{(X^+)^{\Delta}}
\end{align}
\item Stripping off the numerical factor $2(\Delta-2)$ and taking the limit $\Delta\to 2$ for the remaining part, we obtain a new non-pure-gauge mode function that is antisymmetric in $i, j$
\begin{align}\label{gamma1}
\gamma^{(1)}_{ij}(X,U)\equiv \frac{x_i\, u_j - u_i\, x_j}{(X^+)^2}=\frac{1}{2} \left(P_i \frac{u_j}{X^+}-P_j \frac{u_i}{X^+}\right)
\end{align}
\end{itemize}
It's straightforward to check that $\gamma^{(1)}_{ij}$ satisfies all the requirements of being a quasinormal mode of frequency $i\omega=2$. Written in the form of  (\ref{gamma1}), $\gamma^{(1)}_{ij}$ looks like a descendant of $\frac{u_i}{X^+}$. However, this descendant structure doesn't have any physical meaning because $\frac{u_i}{X^+}$ fails to satisfy the tangentiality condition (\ref{tan}). Actually, $\gamma^{(1)}_{ij}$ is a primary up to  gauge transformation:
\begin{align}
K_k\gamma^{(1)}_{ij}=U\cdot\partial_X\left(\frac{\delta_{ik}x_j-\delta_{jk}x_i}{X^+}\right)
\end{align}
Therefore,  $\gamma^{(1)}_{ij}$ is a physical primary quasinormal mode and we will call the whole Verma module built from $\gamma^{(1)}_{ij}$  the ``$\gamma$-tower'' of quasinormal modes. 

In the framework of pseudo gauge symmetry, $\gamma^{(1)}_{ij}$ can be thought as a $U(1)$ field strength with $\frac{u_i}{X^+}$ being the gauge potential. As a field strength, $\gamma^{(1)}_{ij}$ satisfies Bianchi identity $P_{[i}\gamma^{(1)}_{jk]}=0$ which imposes a nontrivial constraint on the descendants of $\gamma^{(1)}_{ij}$. When $d=3$, the field strength $\gamma^{(1)}_{ij}$ is dual to a vector $\tilde\gamma^{(1)}_{i}$ and the Bianchi identity is equivalent to a conservation equation $P_i\tilde\gamma^{(1)}_i=0$. In this case, the representation structure of the $\gamma$-tower is exactly the same as the $\alpha$-tower.

We'll leave the comparison with intrinsic coordinate computation of quasinormal modes of Maxwell theory to appendix \ref{Max2}. 

\subsubsection{Linearized gravity}
$\mathbf{d\ge 4}$

\noindent{}The primary $\beta$-mode associated to a massless spin-2 field is $\beta^{(2)}=\frac{\Phi^2}{(X^+)^2}$. According to eq. (\ref{defxi}) and (\ref{defeta}), $\beta^{(2)}$ can be alternatively expressed as 
\begin{align}
\beta^{(2)}_{ij}= P_i \,\left( U\cdot\partial_X \,\eta^j_1\right)+P_j \,\left( U\cdot\partial_X\, \eta^i_1\right)-\text{trace}
\end{align}
where the null vectors $z$ in $\beta^{(2)}$ are stripped off. Treating $P_i$ as an ordinary derivative, the first two terms in $\beta^{(2)}_{ij}$ have the form of diffeomorphism transformation of (Euclidean) linearized gravity in $\mathbb{R}^d$. Given this pseudo diffeomorphism structure, we can naturally kill these two terms by considering the (pseudo) linearized Riemann tensor $R[\beta^{(2)}]_{ijk\ell}$ , which is defined as 
\begin{align}
R[\beta^{(2)}]_{ijk\ell}\equiv\frac{1}{2}\left(P_{j}P_k\beta^{(2)}_{i\ell}+P_iP_\ell\beta^{(2)}_{jk}-P_{i}P_k \beta^{(2)}_{j\ell}-P_jP_\ell \beta^{(2)}_{ik}\right)
\end{align}
The remaining pure trace term in $\beta^{(2)}_{ij}$ drops out by projecting $R[\beta^{(2)}]_{ijk\ell}$ to the ``linearized Weyl tensor'' $\CC[\beta^{(2)}]_{ijk\ell}$ which is defined as the traceless part $R[\beta^{(2)}]_{ijk\ell}$ and  carries the $\tiny \yng(2,2)$ representation of $\SO(d)$. By construction, $\CC[\beta^{(2)}]_{ijk\ell}$ would vanish  just as the $U(1)$ field strength $\CF_{ij}$ in the spin-1 case. Due to this similarity, it's quite natural to expect the new spin-2 physical primary quasinormal mode will be produced by the  same ``deformation+ limiting'' procedure, whose steps are listed here again for  readers' convenience: $(a)$ deform the $\beta^{(2)}$ mode by sending it to  $\hat\beta^{(2)}\equiv \frac{\Phi^2}{(X^+)^\Delta}=(X^+)^{2-\Delta}\beta^{(2)}$, $(b)$ compute the Weyl tensor $\CC[\hat\beta^{(2)}]_{ijk\ell}$ associated to $\hat\beta^{(2)}$, $(c)$ strip off the overall factor $(\Delta\!-\!2)$  in $\CC[\hat\beta^{(2)}]_{ijk\ell}$ and take the limit $\Delta\to 2$ for the remaining part. To show the limiting procedure $(c)$ more explicitly, expand, for instance, the first term in $R[\hat\beta^{(2)}]$: 
\begin{align}\label{firstR}
P_jP_k \hat \beta^{(2)}_{i\ell}=2(\Delta-2)\,\beta^{(2)}_{i\ell}\,P_j  \frac{x_k}{(X^+)^{\Delta-1}}+2(\Delta-2)\frac{x_{(k}\,P_{j)}\beta^{(2)}_{i\ell}}{(X^+)^{\Delta-1}} + \frac{P_j P_k \beta^{(2)}_{i\ell}}{(X^+)^{\Delta-2}}
\end{align}
where the convention for symmetrization is $x_{(k}\,P_{j)}=x_k\, P_j+x_j \,P_k$. The last term in (\ref{firstR}) does not contribute to the Riemann tensor $R[\hat \beta^{(2)}]$ and we'll drop it henceforth. The remaining terms are proportional to $\Delta\!-\!2$, so the limiting  procedure is applicable to them 
\begin{align}
\lim_{\Delta\to 2}\frac{P_j P_k \hat \beta^{(2)}_{i\ell}}{\Delta-2} &=  P_j  \left(\frac{2\, x_k}{X^+}\right) \beta^{(2)}_{i\ell}+ \frac{2\,x_k}{X^+}P_j\beta^{(2)}_{i\ell}+ \frac{2\, x_j}{X^+}P_k\beta^{(2)}_{i\ell}\nonumber\\
&=P_jP_k(-\log(X^+)\,\beta^{(2)}_{i\ell})+\log(X^+)P_jP_k \,\beta^{(2)}_{i\ell}
\end{align}
where the last term drops out from Riemann tensor  $R[\hat \beta^{(2)}]$. Therefore the new physical quasinormal mode is schematically
\begin{align}\label{gammaC}
\gamma^{(2)}_{ijk\ell}(X, U)\equiv \CC[h]_{ijk\ell}, \,\,\,\, h_{ij}=-\log (X^+) \,\beta^{(2)}_{ij}
\end{align}
which has scaling dimension 2 (or quasinormal frequency $i\omega=2$) and carries the $\tiny\yng(2,2)$ representation of $\SO(d)$. 
The primariness of $\gamma^{(2)}_{ijk\ell}$ is proved in appendix \ref{primariness}.

\vspace{10pt}

\noindent{}$\mathbf{d=3}$

The $d=3$ case is degenerate and requires a separate discussion because the 3D Weyl tensor $\CC[\hat\beta^{(2)}]_{ijk\ell}$ vanishes identically for arbitrary choice of $\Delta$. So the deformation and limiting procedure used above fails to yield any quasinormal mode when $d=3$. The solution to this problem is using Cotton tensor, the 3D analogue of Weyl tensor.  On a 3-dimensional Riemann manifold with metric $g_{ij}$, the Cotton tensor is given by \cite{Cotton:original, Cotton:1}
\begin{align}
\CC^j_i[g] =\nabla_k\left(R_{i\ell}-\frac{1}{4}R\, g_{i\ell}\right) \epsilon^{k\ell j}
\end{align}
and the vanishing of  Cotton tensor is the necessary and sufficient condition for the 3-dimensional manifold to be conformally flat. (In spite of the abuse of notation $\CC$, it will be clear from the context that  $\CC$ means Weyl tensor when $d\ge 4$ and means Cotton tensor when $d=3$). At the linearized level, i.e. $g_{ij}=\delta_{ij}+\phi_{ij}$, the Cotton tensor becomes
\begin{align}
\CC[\phi]_{ij}=\partial_k R_{i \ell}[\phi]\epsilon_{k\ell j}-\frac{1}{4}\epsilon_{kij}\partial_k R[\phi] 
\end{align} 
where $R[\phi]_{ij}, R[\phi]$ are linearized Ricci tensor and linearized Ricci scalar respectively. In addition, the linearized Cotton tensor is actually symmetric in $i, j$ which can be checked by contracting it with $\epsilon_{ijm}$
\begin{align}
\CC[\phi]_{ij}\epsilon_{ijm}=\frac{1}{2}\partial_m R[\phi]-\partial_i R_{im} =0
\end{align}
where the last step is a well-known result of Bianchi identity of Riemann tensor. Since $\CC[\phi]_{ij}$ is symmetric and traceless, it will be convenient to restore the null vector $z$
\begin{align}\label{Cotton}
\CC[\phi; z]=\partial_k R_{i \ell}[\phi]\epsilon_{k\ell j}\, z_i\, z_j=\frac{1}{2}\epsilon_{k\ell j}(\partial_i\partial_k \partial_m\phi_{m\ell}-\partial^2\partial_k\phi_{i\ell})\, z_i\, z_j
\end{align}
In the context of quasinormal modes, we can similarly construct a pseudo Cotton tensor with ordinary derivative $\partial_i$ in eq. (\ref{Cotton}) replaced by momentum operator $P_i$
\begin{align}\label{Cotton1}
\CC[\phi; z]=\frac{1}{2}\epsilon_{k\ell j}(P_iP_k P_m\phi_{m\ell}-P^2P_k\phi_{i\ell})\, z_i\, z_j
\end{align}
 Because Cotton tensor is invariant under diffeomorphism and local Weyl transformation by construction, $\CC[\phi; z]$ vanishes exactly when $\phi=\beta^{(2)}$. As a result,  
applying the deformation and limiting procedure to the Cotton tensor $\CC[\beta^{(2)}; z]$ would yield a new physical primary quasinormal mode
\begin{align}
\gamma^{(2)}(X, U; z)=\frac{z\cdot(x\wedge u)}{(X^+)^3}(X^+ U^- \, x\cdot z- X^+X^-\, u\cdot z+x^2 \, u\cdot z- (u\cdot x)(x\cdot z))
\end{align}
where $(x\wedge u)_i =\epsilon_{ijk}x_j u_k$. Compared to higher dimensional cases, $\gamma^{(2)}$ in $d=3$ is different mainly in two ways: $(a)$ it has scaling dimension 3 because Cotton tensor involves three derivatives while Weyl tensor only involves 2, $(b)$ it carries a spin-2 representation of $\SO(3)$ and the Bianchi identity in higher dimension becomes a ``conservation'' equation $P_i \gamma^{(2)}_{ij}=0$. Due to these two properties, the $\gamma^{(2)}$-tower of quasinormal modes in $\text{dS}_4$ is isomorphic to the $\alpha^{(2)}$-tower.

\subsubsection{Massless higher spin fields}
With the spin-1 and spin-2 examples worked out explicitly, we'll continue to show that for any massless higher spin field, there exist the primary $\gamma$-mode. For a massless spin-$s$ field, the primary $\beta$-mode given by eq. (\ref{defxi}) and (\ref{defeta}) can be written as
\begin{align}
\beta^{(s)}_{i_1\cdots i_s}= P^{(i_1} U\cdot\partial_X \eta_{s-1}^{i_2\cdots i_s)}-\text{trace}
\end{align}
Treating $P^i$ as an ordinary derivative, apart from the pure trace part, $\beta^{(s)}$ has the form of gauge transformation of $d$-dimensional linearized spin-$s$ gravity. Such gauge transformation can be eliminated by using the higher spin Riemann tensor
\begin{align}
R[\phi]_{i_1 \ell_1,\cdots, i_s \ell_s}\equiv \Pi_{ss} P_{i_1}\cdots P_{i_s} \phi_{\ell_1\cdots \ell_s}
\end{align} 
where $\Pi_{ss}$ is a projection operator ensuring $R[\phi]_{i_1 \ell_1,\cdots, i_s \ell_s}$ carries the $\mY_{ss}$ representation of $\text{GL}(d, \mathbb R)$. ($\mY_{nm}$ denotes a 2-row Young diagram with $n$ boxes in the first row and $m$ boxes in the second. When $m=0$, we use $\mY_n$.) More explicitly, $\Pi_{ss}$ can be realized by antisymmetrizing the $s$ pairs of indices: $[i_1, \ell_1], \cdots [i_s, \ell_s]$ \cite{Bekaert:2002dt, TD}. The higher spin Weyl tensor $\CC[\phi]_{i_1 \ell_1,\cdots, i_s \ell_s}$ is defined as the traceless part of $R[\phi]_{i_1 \ell_1,\cdots ,i_s \ell_s}$ and thus it carries the $\mY_{ss}$ representation of $\SO(d)$ \footnote{For simplicity, we assume $d\ge 4$ so the higher spin Weyl tensor is nonvanishing. When $d=3$, we should use higher spin Cotton tensor $\CC_{i_1\cdots i_s}$ \cite{TD, Henneaux:2015cda} that is symmetric and traceless. The Bianchi identity for Cotton tensor is  a conservation equation $P_{i_1} \CC_{i_1\cdots i_s}=0$. In addition, the definition of Cotton tensor involves $2s-1$ momentum operators and hence the associated primary $\gamma$-mode would have scaling dimension $1+s$ instead of 2.}. Since Weyl tensor is invariant under diffeomorphism and local Weyl transformation, $\CC[\beta^{(s)}]$ vanishes exactly. Thus we can apply the deformation procedure to it and obtain the following  quasinormal mode
\begin{align}\label{gammas}
\boxed{\gamma^{(s)}_{i_1 \ell_1,\cdots i_s \ell_s}(X, U)\equiv \CC[h^{(s)}]_{i_1 \ell_1,\cdots, i_s \ell_s}, \,\,\,\,h^{(s)}_{\ell_1\cdots \ell_s}=-\log(X^+)\beta^{(s)}_{\ell_1\cdots\ell_s}}
\end{align} 
In appendix \ref{primariness}, we show that $\gamma^{(s)}_{i_1 \ell_1,\cdots, i_s \ell_s}$ represents $D^d_{ss}$ primary quasinormal modes of scaling dimension 2 that carry $\mY_{ss}$ representation of $\SO(d)$, where $D^d_{ss}$ is the dimension of the $\mY_{ss}$ representation. (We'll use $D^d_{nm}$ for the dimension of $\mY_{nm}$ representation of $\SO(d)$ and $D^d_{n}$ for the dimension of $\mY_{n}$ representation.)  In the same appendix, we also show that these primary quasinormal modes can alternatively be expressed in the following form \begin{align}\label{exam}\boxed{\gamma^{(s)}(X, U)=\frac{T_{ss}(x, u)}{(X^+)^2}}\end{align}
where $T_{ss}(x, u)$ is a homogeneous polynomial in both $x$ and $u$ of degree $s$ and satisfies (\ref{DefineTSS}). The space of  such $T_{ss}$ carries the $\mY_{ss}$ representation of $\SO(d)$ \cite{Didenko:2014dwa}. One obvious example of $T_{ss}$ is
\begin{align}\label{lwex}
T_{ss}(x,u)=\left[(x^1+ix^2)(u^3+iu^4)-(x^3+ix^4)(u^1+i u^2)\right]^s
\end{align} 
In the representation language, the example given by eq. (\ref{lwex}) is actually the lowest-weight state in $\mY_{ss}$. Therefore, we are able to generate the whole $\gamma$-tower by acting $P_i$ and $M_{ij}$ on the following quansinormal mode:
\begin{align}\label{lw}
\boxed{\gamma_{lw}^{(s)}(X, U)=\frac{\left[(X^1+iX^2)(U^3+iU^4)-(X^3+iX^4)(U^1+i U^2)\right]^s}{(X^+)^2}}
\end{align} 
This is a strikingly universal expression that works for any $s\ge 1$ and $d\ge 4$. In static patch coordinate, the nonvanishing components of (\ref{lw}) are 
\small
\begin{align}\label{lw1}
\gamma^{(s)}_{lw, a_1\cdots a_s}(t,r,\Omega)=\frac{r^{2s}e^{-2\,t}}{(1-r^2)}(\Omega_{12}\partial_{\vartheta^{a_1}}\Omega_{34}-\Omega_{34}\partial_{\vartheta^{a_1}}\Omega_{12})\cdots (\Omega_{12}\partial_{\vartheta^{a_s}}\Omega_{34}-\Omega_{34}\partial_{\vartheta^{a_s}}\Omega_{12})
\end{align}
\normalsize
where $\Omega_{12}=\Omega_1+i\Omega_2$ and $\Omega_{34}=\Omega_3+i\Omega_4$.  One can check that the $\Omega$-dependent part of (\ref{lw1}) is actually a divergence-free spin-$s$ tensor harmonics on $S^{d-1}$. This is also expected from the representation side because these tensor harmonics also furnish the $\mY_{ss}$ representation of $\SO(d)$. 

For the completeness of the final result, we also give the unique lowest-weight state in the $\alpha^{(s)}$-tower here
\begin{align}\label{lw2}
\boxed{\alpha_{lw}^{(s)}(X, U)=\frac{\left[X^+(U^1+i \,U^2)-U^+ (X^1+i \,X^2)\right]^s}{(X^+)^2}\left(\frac{R}{X^+}\right)^{d+2(s-2)}}
\end{align} 
Then all the quasinormal modes are built from $\alpha_{lw}^{(s)}$ (cf. (\ref{lw2})) and $\gamma_{lw}^{(s)}$ (cf. (\ref{lw})) with the action of $P_i$ and $M_{ij}$.

\section{Quasinormal modes from a QFT point of view}\label{southernpole}
In the previous section, we presented a pure algebraic method to construct quasinormal modes of scalars and  higher spin fields. In this section, we'll provide a simple physical picture for this method from a bulk QFT point of view. In particular, we'll use scalar fields and Maxwell fields to illustrate this intuitive picture explicitly and then give a brief comment on general massless higher spin fields. Through out this section, the bulk quantum fields are defined in the southern past planar coordinate $(\eta, y^i)$ of $\text{dS}_{d+1}$ (the region ``S''+``P'' in fig. (\ref{QFTPOV})):
\begin{align}\label{planar}
X^0=\frac{1+y^2-\eta^2}{2\eta},\,\,\,\,\,  X^i=-\frac{y^i}{\eta},\,\,\,\,\, X^{d+1}=-\frac{1-y^2+\eta^2}{2\eta}
\end{align}
where $\eta<0$ and the quasinormal modes are still defined in the southern static patch, which corresponds to  $y<-\eta$ in eq. (\ref{planar}).

\subsection{Scalar fields}
Let $\varphi$ be a scalar field of scaling dimension $\Delta$. Near the past boundary, it has the following asymptotic behavior
\begin{align}\label{asym}
\varphi(\eta,y)\approx (-\eta)^\Delta \CO^{(\alpha)}(y)+ (-\eta)^{\bar\Delta} \CO^{(\beta)}(y)
\end{align}
Define quantum operators $\mathcal{L}_{AB}$ such that $L_{AB}\varphi=-[\mathcal{L}_{AB}, \varphi]$. Then $\CO^{(\alpha)}$ and $\CO^{(\beta)}$ are primary operators in the sense that 
\begin{align}
&[\mathcal{D}, \CO^{(\alpha)}(0)]=\Delta \CO^{(\alpha)}(0),\,\,\,\,\,[\mathcal{K}_i, \CO^{(\alpha)}(0)]=0\nonumber\\
&[\mathcal{D}, \CO^{(\beta)}(0)]=\bar\Delta \CO^{(\beta)}(0),\,\,\,\,\,[\mathcal{K}_i, \CO^{(\beta)}(0)]=0
\end{align}
The bulk two-point function of $\varphi$ defined with respect to the Euclidean vacuum $|E\rangle$ is given by \cite{Bousso:2001mw}
\begin{align}\label{EE}
\langle E|\varphi(X)\varphi(X')|E\rangle=\frac{\Gamma(\Delta)\Gamma(\bar\Delta)}{(4\pi)^{\frac{d+1}{2}}\Gamma(\frac{d+1}{2})} F\left(\Delta,\bar\Delta,\frac{d+1}{2},\frac{1+P}{2}\right)
\end{align}
where $P=X\cdot X'$. We push $X'$ to the past southern pole, i.e. $y'^i=0$ and $\eta'\to 0^-$, then $P$ is approximately $-\frac{X^+}{2\eta'}\to\infty$. For $P\to \infty$, the hypergeometric function in (\ref{EE}) has two leading asymptotic behaviors: $P^{-\Delta}$ and $P^{-\bar\Delta}$. Schematically, it means 
\begin{align}\label{EE1}
\langle E|\varphi(X)\varphi(\eta'\to 0, y'^i=0)|E\rangle\approx c_{\Delta} \frac{(-\eta')^\Delta}{(X^+)^\Delta}+ c_{\bar\Delta} \frac{(-\eta')^{\bar\Delta}}{(X^+)^{\bar\Delta}}
\end{align}
where $c_\Delta$ and $c_{\bar\Delta}$ are two constants. Comparing the eq. (\ref{asym}) and eq. (\ref{EE1}), we find that $\langle E|\varphi(X)\CO^{(\alpha)}(0)|E\rangle$ produces the primary quasinormal mode $\alpha_\Delta(X)$ and $\langle E|\varphi(X)\CO^{(\beta)}(0)|E\rangle$ produces the primary quasinormal mode $\beta_\Delta(X)$. Altogether, the scalar primary quasinormal modes in southern static patch can be produced by inserting primary operator $\CO^{(\alpha)}$ or $\CO^{(\beta)}$ at the southern pole of the past sphere and other quasinormal modes can be produced by inserting descendants of  $\CO^{(\alpha)}$ or $\CO^{(\beta)}$.

\subsection{Maxwell fields}\label{Max}
We want to derive the two primary quasinormal modes of Maxwell field in $\text{dS}_4$ using local operators. First, let's pull back $\alpha^{(1)}_i$ and $\gamma^{(1)}_{ij}$ to planar patch coordinates.

\noindent{}\textbf{The $\alpha$-mode}:
\begin{align}
\alpha^{(1)}_{i,\eta}=\frac{2\, y_i\,\eta^2}{(\eta^2-y^2)^3},\,\,\,\,\, \alpha^{(1)}_{i,j}=-\frac{\eta(2 \,y_i \,y_j+\delta_{ij}\,(\eta^2-y^2))}{(\eta^2-y^2)^3}
\end{align}
For later convenience, we perform a gauge transformation to kill the timelike component, which can be done by choosing the following gauge parameter \footnote{Since quasinormal modes are still defined in the static patch, which corresponds to $\eta+y<0$, we are away from the branch cut of logarithm and the gauge parameter is real.}
\begin{align}
\xi=\frac{y_i}{4\, y^3}\left(\frac{y\,\eta(\eta^2+y^2)}{(\eta^2-y^2)^2}-\frac{1}{2}\log\frac{\eta+y}{\eta-y}\right)
\end{align}
The resulting  spatial part of $\alpha^{(1)}_i$ becomes
\begin{align}\label{gaugealpha}
\boxed{\tilde \alpha^{(1)}_{i, j}=\left(\partial_{y^i}\partial_{y^j}-\delta_{ij}\partial_y^2\right)\frac{\log\frac{\eta+y}{\eta-y}}{8\,y}}
\end{align}

\noindent{}\textbf{The $\gamma$-mode}:
\begin{align}\label{gammalocal}
\boxed{\gamma^{(1)}_{ij, \eta}=0 ,\,\,\,\,\, \gamma^{(1)}_{ij, k}=\frac{y_i\, \delta_{jk}-y_j\,\delta_{ik}}{(\eta^2-y^2)^2}}
\end{align}
The timelike component is automatically vanishing.

Next, we do a mode expansion for a Maxwell field $A_\mu$ in the Coulomb gauge \cite{Anninos:2017eib}:
\begin{align}
 A_i (\eta, y)=-\int\, \frac{d^3 k}{(2\pi)^3}\left(\CO^{(\alpha)}_i(k)\,\frac{\sin(k\eta)}{k}-\CO^{(\beta)}_i(k)\cos(k\eta)\right)e^{\, ik\cdot y}
\end{align}
where the two primary operators $\CO^{(\alpha)}_i, \CO^{(\beta)}_i$ capture the leading asymptotic behavior of $A_i$ near the past boundary
\begin{align}
A_i(\eta\to 0^-, y)\approx  (-\eta) \, \CO_i^{(\alpha)}(y)+ \CO_i^{(\beta)}(y)
\end{align}
They also satisfy the following vacuum two-point functions in momentum space:
\begin{align}
&\langle E|\CO^{(\beta)}_i(k)\CO^{(\beta)}_j(k')|E\rangle=\frac{1}{2k}\left(\delta_{ij}-\frac{k_ik_j}{k^2}\right)(2\pi)^3\delta^3(k+k')\nonumber\\
&\langle E|\CO^{(\alpha)}_i(k)\CO^{(\beta)}_j(k')|E\rangle=\frac{i}{2}\left(\delta_{ij}-\frac{k_ik_j}{k^2}\right)(2\pi)^3\delta^3(k+k')\nonumber\\
&\langle E|\CO^{(\alpha)}_i(k)\CO^{(\alpha)}_j(k')|E\rangle=\frac{k}{2}\left(\delta_{ij}-\frac{k_ik_j}{k^2}\right)(2\pi)^3\delta^3(k+k')\nonumber\\
&\langle E|\CO^{(\beta)}_i(k)\CO^{(\alpha)}_j(k')|E\rangle=-\frac{i}{2}\left(\delta_{ij}-\frac{k_ik_j}{k^2}\right)(2\pi)^3\delta^3(k+k')
\end{align}
Like in the scalar case, we insert the primary operator $\CO^{(\alpha)}_i$ at the southern pole of the past sphere and it produces a mode in the bulk:
\begin{align}\label{Aa}
\langle E | A_i(\eta, y)\CO_j^{(\alpha)}(0)|E\rangle&=-\frac{i}{2}\int\, \frac{d^3 k}{(2\pi)^3}\left(\delta_{ij}-\frac{k_i k_j}{k^2}\right) e^{i k\cdot y-ik \eta}\nonumber\\
&=-\frac{i}{4\pi^2}(\partial_{y^i}\partial_{y^j}-\delta_{ij}\partial_y^2)\int_0^\infty \frac{dk}{k \,y}\sin (k) e^{-i\, \frac{\eta}{y}k}
\end{align}
where the integral over $k$ depends on the relative size of $y$ and $-\eta$ because \footnote{The two cases can be uniformly treated if we give $a$ a small positive imaginary part, i.e. $a\to a+i\epsilon, \epsilon>0$. With this $i\epsilon$ prescription,  $\frac{i}{2}\log\frac{a+i\epsilon+1}{a+i\epsilon-1}$ works for both cases. In bulk, it amounts to Wick rotating the planar coordinate time: $\eta\to e^{i\epsilon}\eta$.}
\begin{align}\label{int}
\int_0^\infty\frac{dk}{k}\sin (k) e^{i\,a\,k}=\begin{cases}\frac{i}{2}\log\frac{a+1}{a-1},\,\,\,\,\,&|a|>1 \\ \frac{\pi}{2}+\frac{i}{2}\log\frac{1+a}{1-a}, \,\,\,\,\, & |a|<1\end{cases}
\end{align}
In a physical picture, the jumping at $|a|=1$ reflects horizon crossing.
For quasinormal modes defined in southern static patch, which has $y<-\eta$, the $k$-integral in (\ref{Aa}) corresponds to the top case of (\ref{int}):
\begin{align}
\langle E | A_i(\eta, y)\CO_j^{(\alpha)}(0)|E\rangle=(\partial_{y^i}\partial_{y^j}-\delta_{ij}\partial_y^2)\left(\frac{-1}{8\pi^2 y}\log\frac{\eta+y}{\eta-y}\right)
\end{align}
Up to normalization, we precisely reproduce the $\alpha^{(1)}_{i}$ quasinormal mode given by (\ref{gaugealpha}). Similarly, we can insert the $\CO^{(\beta)}_i$ at the southern pole of the past sphere and it yields
\begin{align}\label{Abeta}
\langle E|A_k(\eta, y)\CO^{(\beta)}_i(0)|E\rangle=\, \int \frac{d^3 k}{(2\pi)^3}\, \left(\delta_{ik}-\frac{k_i k_k}{k^2}\right)\frac{e^{ik\cdot y-ik\eta}}{2k}
\end{align}
Due to the extra $\frac{1}{k}$ in the integrand compared to eq.(\ref{Aa}), this mode suffers from an IR divergence around $k=0$. However, this divergence can be eliminated if we replace $\CO^{(\beta)}_{i}(0)$ by a ``curvature'' $\CO^{(\gamma)}_{ij}(0)\equiv P_i\CO^{(\beta)}_j(0)-P_j\CO^{(\beta)}_i(0)$ as in the construction of $\gamma^{(1)}_{ij}$. The insertion of $\CO^{(\gamma)}_i$ at southern pole yields
\small
\begin{align}\label{gamma}
\langle E|A_k(\eta, y)\CO^{(\gamma)}_{ij}(0)|E\rangle=\frac{i}{2}\int\, \frac{d^3 k}{(2\pi)^3}\,\frac{k_i\delta_{jk}-k_j \delta_{ik}}{k}\, e^{ik\cdot y-ik\eta}=\frac{1}{2\pi^2}\frac{y_j\delta_{ik}-y_i\delta_{jk}}{(\eta^2-y^2)^2}
\end{align}
\normalsize
which is exactly the $\gamma^{(1)}_{ij}$ quasinormal mode, cf.  (\ref{gammalocal}), up to normalization. Note that in the definition of $\CO^{(\gamma)}_{ij}$, we implicitly use the {\it pseudo} gauge symmetry structure. On the other hand, $P_i$ reduces to ordinary derivative $\partial_i$ at boundary and hence $\CO^{(\gamma)}_{ij}$ is indeed the curvature corresponding to the  boundary gauge symmetry. Therefore, the classically pseudo gauge symmetry can be identified as the boundary gauge symmetry in quantum theory.
In this sense, $\CO^{(\gamma)}_{ij}\sim \epsilon_{ijk} B_k$ has the interpretation as  a boundary magnetic field and $\CO^{(\alpha)}_i\sim E_i$ has the interpretation as a boundary electric field, subject to the constraint $\nabla\cdot E=0$. So the quasinormal modes of Maxwell fields are produced by the electric/magnetic field operator, together with their derivatives, inserted at the past southern pole, cf. fig (\ref{QFTPOV}).

In general, a free massless higher spin field $\varphi_{\mu_1\cdots \mu_s}$ in a suitable gauge has the following asymptotic behavior near the past boundary
\begin{align}
\varphi_{i_1\cdots i_s}(\eta\to 0^-, y)\approx (-\eta)^{d-2}\CO^{(\alpha)}_{i_1\cdots i_s}(y)+ (-\eta)^{2-2s}\CO^{(\beta)}_{i_1\cdots i_s}(y)
\end{align}
where $\CO^{(\alpha)}_{i_1\cdots i_s}(y)$ is a gauge-invariant boundary conserved current and $\CO^{(\beta)}_{i_1\cdots i_s}(y)$ is a boundary gauge field. \footnote{For a bulk gauge transformation $\delta\varphi_{\mu_1\cdots\mu_s}=\nabla_{(\mu_1}\xi_{\mu_2\cdots\mu_s)}$, the asymptotic behavior of $\xi$ near the past boundary is $\xi_{i_1\cdots i_s}(\eta, y)\approx (-\eta)^{d}\, A_{i_1\cdots i_{s-1}}(y)+ (-\eta)^{2-2s}B_{i_1\cdots i_{s-1}}(y)$. $\CO^{(\alpha)}_{i_1\cdots i_s}$ is clearly invariant under this transformation as the $A$-mode falls off too fast to affect it. Meanwhile $\CO^{(\beta)}_{i_1\cdots i_s}$ undergoes an induced boundary gauge transformation $\delta\CO^{(\beta)}_{i_1\cdots i_s}=\partial_{(i_1} B_{i_2\cdots i_s)}$ because the $B$-mode has the same fall-off as it.} From $\CO^{(\beta)}_{i_1\cdots i_s}(y)$, we can build a boundary Weyl tensor $\CO^{(\gamma)}_{i_1j_1,\cdots, i_sj_s}$ that is gauge invariant. Then inserting  operators in the conformal family of $\CO^{(\alpha)}_{i_1\cdots i_s}$ at the past southern pole produces the $\alpha$-tower of quasinormal modes and inserting operators in the conformal family of $\CO^{(\gamma)}_{i_1j_1,\cdots, i_sj_s}$ at the past southern pole produces the $\gamma$-tower of quasinormal modes

\section{Quasinormal modes and $\SO(1, d+1)$ characters}\label{char}
In the section \ref{construction}, we describe a procedure to construct quasinormal modes of scalar fields and higher spin fields in $\text{dS}_{d+1}$. In this section, we'll extract the whole quasinormal spectrum by using this construction and show that it's related to the Harish-Chandra group character of $\SO(1, d+1)$. To collect the information of quasinormal modes of certain field $\phi$ in a compact expression, we define a ``quasinormal character'' :
\begin{align}\label{quacha}
\chi^{\text{\text{QN}}}_\phi(q)=\sum_{\omega} \, d_\omega \,q^{i\omega}, \,\,\,\,\, 0<q<1
\end{align}
where the sum runs over all quasinormal frequencies of $\phi$ and $d_\omega$ is the degeneracy of quasinormal modes with frequency $\omega$. Due to the representation structure of the quasinormal modes, the quasinormal character $\chi^{\text{QN}}_\phi(q)$ naturally splits into two different parts, with each part involves either the $\alpha$-tower or $\beta/\gamma$-tower of quasinormal modes. For example, let $\phi$ be a real scalar field of scaling dimension $\Delta$. The quasinormal modes in the $\alpha$-tower have frequencies $i\omega=\Delta+n, n\ge 0$ and for each $n$ the degeneracy is $\binom{d+n-1}{d-1}$. Thus the $\alpha$-part of the  quasinormal character is 
\begin{align}
\chi^{\text{QN}, \alpha}_\Delta(q)=\sum_{n\ge 0}\binom{d+n-1}{d-1} \, q^{\Delta+n}=\frac{q^\Delta}{(1-q)^d}
\end{align}
Similarly, the contribution of $\beta$-tower is $\chi^{\text{QN}, \beta}_\Delta(q)=\frac{q^{\bar\Delta}}{(1-q)^d}$. Altogether, we obtain the full quasinormal character of $\phi$
\begin{align}
\chi^{\text{QN}}_\Delta(q)=\chi^{\text{QN}, \alpha}_\Delta(q)+\chi^{\text{QN}, \beta}_\Delta(q)=\frac{q^\Delta+q^{\bar \Delta}}{(1-q)^d}
\end{align}
According to \cite{Basile:2016aen}, $\frac{q^\Delta+q^{\bar \Delta}}{(1-q)^d}$ is exactly the Harish-Chandra character $\chi^{\text{HC}}_\Delta(q)\equiv \tr q^D$ \footnote{In our convention, $D$ is an anti-hermitian operator in unitary representations.} for the scalar principal series, i.e. $\Delta\in \frac{d}{2}+i\,\mathbb R$ and  the scalar complementary series, i.e. $0<\Delta<d$. Note that the principal/complementary series condition just ensures a positive mass term for $\phi$, which in bulk is nothing but the unitarity condition. Therefore, the quasinormal character of a unitary scalar field is same as its Harish-Chandra character. For a massive spin-$s$ field, the story is almost the same except the $\alpha$-modes and $\beta$-modes have spin degeneracy $D^d_s$,  the dimension of spin-$s$ representation of $\SO(d)$. Taking into account this spin degeneracy, we obtain the quasinormal character of a spin-$s$ field of scaling dimension $\Delta$,
\begin{align}\label{quachar1}
\chi^{\text{QN}}_{[\Delta,s]}(q)=D^d_s\, \frac{q^\Delta+q^{\bar \Delta}}{(1-q)^d}
\end{align}
which is exactly the Harish-Chandra character for the  spin-$s$ principal series, i.e. $\Delta\in \frac{d}{2}+i\,\mathbb R$ and  the spin-$s$ complementary series, i.e. $1<\Delta<d-1$.

 In the remaining part of this section, we'll compute quasinormal characters for massless higher spin fields. In this case,  the $\alpha$-part  is easy because $\alpha^{(s)}_{i_1\cdots i_s}$ is a conserved current in the sense of (\ref{PCD}). So for $i\omega^\alpha_n=d+s-2+n$ in the $\alpha$-tower, the degeneracy  is $d_n^\alpha=\binom{n+d-1}{d-1}D^d_s-\binom{n+d-2}{d-1}D^d_{s-1}$, which yields
\begin{align}\label{QNs}
\chi^{\text{QN}, \alpha}_s(q)=\sum_{n\ge 0}d^\alpha_n\,  q^{i\omega^\alpha_n}=\frac{D^d_s\, q^{d-2+s}-D^d_{s-1}\,q^{d-1+s}}{(1-q)^d}
\end{align}
$\chi^{\text{QN}, \alpha}_s(q)$ is the same as the $\SO(2,d)$ character corresponding to a massless spin-$s$ field in $\text{AdS}_{d+1}$\cite{Dolan:2005wy, Gibbons:2006ij}. However, this is far from the corresponding $\SO(1,d+1)$ character \cite{Hirai:Char,Basile:2016aen}. In the notation of \cite{Hirai:Char}, the massless spin-$s$ representation of $\SO(1,d+1)$ is denoted by $D^j_{(\alpha;p)}$, with $p=0, j=\frac{d-4}{2}$ for even $d$, $j=\frac{d-3}{2}$ for odd $d$ and $\alpha=(s,s,0,\cdots,0)$\footnote{Here we use a different convention for the highest weight vector $\alpha$ compared to \cite{Hirai:Char}.}. In the notation of \cite{Basile:2016aen}, this representation falls in the exceptional series with $\Delta=p=2$ and $\mY_p=\mY_{ss}$. The corresponding Harish-Chandra $\SO(1,d+1)$ character is 
\begin{align}\label{ex1}
\chi^{\text{HC}}_s(q)=(1-(-1)^d)\frac{D^d_s\,q^{s+d-2}-D^d_{s-1}\,q^{s+d-1}}{(1-q)^d}+\sum_{n=2}^{d-2}(-)^n\frac{ \CD_n q^n}{(1-q)^d}
\end{align}
where 
\begin{align}
\CD_n=\frac{\Gamma(d-3)s(s+1)(d+s-4)(d+s-3)D^d_{ss}}{\Gamma(n-1)\Gamma(\bar n-1)(s+n-2)(s+\bar n-2)(s+n-1)(s+\bar n-1)},\,\,\,\, \bar n\equiv d-n
\end{align}
For $2\le n\le \floor*{\frac{d}{2}}$, $\CD_n$ is the dimension of $\SO(d)$ representation $\mY_{(ss, 1^m)}$, obtained by adding $n-2$ single-box rows to $\mathbb{Y}_{ss}$. (When $d=2r$ and $n=r$, $\CD_r$ is actually the dimension of $\mY_{(ss1\cdots 1,+1)}\oplus \mY_{(ss1\cdots1,-1)}$). These values can be easily extended to $\floor*{\frac{d}{2}}+1\le n\le d-2$ by the manifest $n\leftrightarrow d-n$ symmetry of $\CD_n$.

To compare the quasinormal character $\chi^{\text{QN}}_s$ with the Harish-Chandra character $\chi^{\text{HC}}_s$, we still need to figure out quasinormal spectrum of the $\gamma$-tower.

\noindent{\textbf{Maxwell field}}

Let's start from a Maxwell field. At level 0, i.e. $i\omega=2$, the degeneracy is $d^\gamma_0=\binom{d}{2}$ because $\gamma^{(1)}_{ij}$ carries the 2-form representation of $\SO(d)$.
At level 1, generic descendants are of the form $P_k \gamma^{(1)}_{ij}$, corresponding to the $\SO(d)$ representation $\tiny{\yng(1)}\otimes \tiny{\yng(1,1)}$. The 3-form representation in this tensor product is vanishing due to Bianchi identity $P_{[k} \gamma^{(1)}_{ij]}=0$. Therefore the degeneracy of quasinormal modes with frequency $i\omega=3$ is 
\begin{align}
d^\gamma_1=d \binom{d}{2}-\binom{d}{3}=2\binom{d+1}{3}
\end{align}
The descendants at level 2 are $P_k P_\ell \gamma^{(1)}_{ij}$, corresponding to the $\SO(d)$ representation $(\bullet\oplus \tiny{\yng(2)})\otimes \tiny{\yng(1,1)}$. Due to Bianchi identity, we would exclude terms like $P_\ell P_{[k} \gamma^{(1)}_{ij]}$, that carries the $\tiny{\yng(1)}\otimes\tiny{\yng(1,1,1)}$ representation. However, this is overcounting because the 4-form representation in this tensor product, carried by $P_{[\ell} P_{k} \gamma^{(1)}_{ij]}$, vanishes automatically without using Bianchi identity. Therefore the degeneracy of quasinormal modes with frequency $i\omega=4$ is 
\begin{align}
d^\gamma_2=\binom{d+1}{d-1} \binom{d}{2}-d \binom{d}{3}+\binom{d}{4}=3\binom{d+2}{4}
\end{align}
At any level $n$, using the same argument, we obtain the degeneracy of quasinormal modes with frequency $i\omega=2+n$ 
\begin{align}
d_n^\gamma=\sum_{k=0}^n(-)^{k}\binom{d}{k+2}\binom{n+d-1-k}{d-1}=(n+1)\binom{n+d}{d-2}
\end{align}
which leads to the $\gamma$-tower quasinormal character
\begin{align}\label{QB}
\chi^{\text{QN},\gamma}_1(q)=\sum_{n\ge 0}(n+1)\binom{n+d}{d-2}q^{n+2}=1-\frac{1-dq}{(1-q)^d}
\end{align}
Combining eq.(\ref{QNs}) for $s=1$ and eq.(\ref{QB}), we get the full quasinormal character of a Maxwell field
\begin{align}\label{QN1}
\chi^{\text{QN}}_1(q)=\chi^{\text{QN},\alpha}_1(q)+\chi^{\text{QN},\gamma}_1(q)=1-\frac{1-dq}{(1-q)^d}+\frac{dq^{d-1}-q^d}{(1-q)^d}
\end{align}
On the other hand, since $\CD_n$ reduces to $\binom{d}{n}$ when $s=1$, the Harish-Chandra character (\ref{ex1}) for $s=1$ is 
\begin{align}
\chi^{\text{HC}}_1(q)=\frac{d\,q^{d-1}-q^{d}}{(1-q)^d}+\sum_{n=2}^{d}(-)^n\frac{ \CD_n q^n}{(1-q)^d}=\chi^{\text{QN}}_1(q)
\end{align}
Again, quasinormal character=Harish-Chandra character.  In appendix \ref{match}, we'll show that $\chi^{\text{QN}}_1(q)$ is also consistent with the spin-1 quasinormal spectrum  in \cite{LopezOrtega:2006my}.

\noindent{\textbf{Higher spin fields}}

To check $\chi^{\text{QN}}_s=\chi^{\text{HC}}_s$ for any $s\ge 2$, it is easier to use a different but equivalent expression of $\chi^{\text{HC}}_s$  found in \cite{Anninos:2020hfj}: 
\begin{align}\label{flipping}
\chi^{\text{HC}}_s(q)=\frac{D^d_s q^{d+s-2}-D^d_{s-1}q^{s+d-1}}{(1-q)^d}+\left[\frac{D^d_s q^{2-s}-D^d_{s-1}q^{1-s}}{(1-q)^d}\right]_+
\end{align} 
where $[\,\,\,\,]_+$ is a linear operator  that sends $q^k\to -q^{-k}$ for $k<0$ and drops the constant term while acting a Laurent series around $q=0$. As a very simple example, $[q^{-1}\!+\!1\!+\!q]_+=0$. Notice that the first term of $\chi^{\text{HC}}_s(q)$ is the same as $\chi^{\text{QN},\alpha}_s(q)$, so it suffices to compare the second term with $\chi^{\text{QN},\gamma}_s(q)$. Expand the second term into a Taylor series around $q=0$:
\begin{align}
\left[\frac{D^d_s q^{2-s}-D^d_{s-1}q^{1-s}}{(1-q)^d}\right]_+\equiv \sum_{n\ge 0} b_n q^{2+n} 
\end{align}
With some simple algebra, one can show that $b_0=D^d_{ss}, b_1=D^d_{s+1,s}+D^{d}_{s,s-1}$ and furthermore $b_n$ satisfies the following recurrence relation 
\begin{align}\label{recb}
b_{n}-b_{n-2}=D^d_s(D^d_{s+n}+D^d_{s-n-2})-D^d_{s-1}(D^d_{s+n+1}+D^d_{s-n-1})
\end{align}
Using the tensor product decomposition of $\mY_s\otimes \mY_t$ (assuming $t\le s$)
\begin{align}
\mY_s\otimes \mY_t=\bigoplus_{\ell=0}^t \bigoplus_{m=0}^{t-\ell}\mY_{s+t-2\ell-m, m}
\end{align}
the products of dimensions in eq. (\ref{recb}) can be rewritten as a summation
\begin{align}
b_{n}-b_{n-2}=\sum_{\ell=0}^s D^d_{s-\ell+n, s-\ell}-\sum_{\ell=0}^{s-n-1} D^d_{s-\ell-1, s-n-1-\ell}=\left(\sum_{\ell=0}^s-\sum_{\ell=n+1}^s\right) \, D^d_{s-\ell+n, s-\ell}
\end{align}
When $n\ge s$ the second sum in the bracket vanishes and when $n\le s$, the first sum over $\ell$ gets truncated at $\ell=n$. Altogether, 
\begin{align}
b_{n}-b_{n-2}=\sum_{\ell=0}^{\text{min}(n, s)}D^d_{s-\ell+n, s-\ell}
\end{align}
On the quasinormal modes side, since the primary  $\gamma^{(s)}$ carries $\mathbb Y_{ss}$  representation, the degeneracy at level 0 is $d^{\gamma}_0=D^d_{ss}=b_0$. At level 1, the descendants $P_k \gamma^{(s)}_{i_1j_1,\cdots, i_s j_s}$ are represented by $\mY_1\otimes \mY_{ss}=\mY_{s+1,s}\oplus \mY_{s,s-1}\oplus\mY_{s,s,1}$. Due to Bianchi identity, the three-row summand in this tensor product vanishes and the level 1 descendants only carry the $\mY_{s+1,s}\oplus \mY_{s,s-1}$ representation. So the degeneracy of quasinormal frequency $i\omega=3$ is $d^{\gamma}_1=D^d_{s+1,s}+D^d_{s,s-1}=b_1$. At higher levels, we aim to derive a recurrence relation for the degeneracy $d_n^\gamma$.  For example, at level $n$, the descendants are of the form $P_{\ell_1}\cdots P_{\ell_n} \gamma^{(s)}_{i_1 j_1,\cdots, i_s j_s}$, where $P_{\ell_1}\cdots P_{\ell_n} $ should be understood group theoretically the symmetrized tensor product of $n$ spin-1 representations. Compared to level $(n-2)$, the new representation structure is $\mY_n\otimes \mY_{ss}$ where $\mY_n$ corresponds to the traceless part of $P_{\ell_1}\cdots P_{\ell_n} $ and $\mY_{ss}$ corresponds to $\gamma^{(s)}$. Due to Bianchi identity, only two-row representations in the tensor product decomposition of $\mY_n\otimes \mY_{ss}$ are nonvanishing. These two-row representations are exactly $\oplus_{\ell=0}^{\text{min}(n, s)}\mY_{s-\ell+n, s-\ell}$, which yields $d_n^\gamma-d^\gamma_{n-2}=b_n-b_{n-2}$ and furthermore $b_n=d^\gamma_n$ for $n\ge 0$. Altogether, we can conclude 
\begin{align}
\boxed{\chi^{\text{QN}}_s(q)=\chi^{\text{HC}}_s(q)}
\end{align}

\section{Conclusion and outlook}\label{finally}
In this paper, we present an algebraic method of constructing quasinormal modes of massless higher spin fields in the southern static patch of $\text{dS}_{d+1}$ using ambient space formalism. With the action of isometry group $\SO(1,d+1)$, the whole quasinormal spectrum can be built from two primary quasinormal modes, whose properties are summarized in the table.\ \ref{table} (assuming $d\ge 4$)

\begin{table}[h]
\centering
\begin{tabular}{ |c|c|c|c| } 
 \hline
 Primaries & $i\,\omega_{\text{QN}}$ & $\SO(d)$ representation & constraint\\ 
 \hline
$\alpha^{\mu_1\cdots\mu_s}_{i_1\cdots i_s}$ & $d+s-2$ & $\mY_s$ & conservation law \\ 
 \hline
$\gamma^{\mu_1\cdots\mu_s}_{i_1j_1,\cdots, i_s j_s}$ & 2 & $\mY_{ss}$ & Bianchi identity \\ 
 \hline
\end{tabular}
\caption{\label{table} A brief summary about the {\it physical} primary quasinormal modes of a massless spin-$s$ gauge field in $\text{dS}_{d+1}$ ($d\ge 4$). $\mu_k$ are bulk spin indices and $i_k, j_k$ indicate the $\SO(d)$ representation whose dimension gives the degeneracy.}
\end{table}


For example, when $s=2$, the primary $\alpha$-modes $\alpha^{\mu\nu}_{i_1 i_2}$ have quasinormal frequency $\omega_{QN}=-id$ and degeneracy $D^d_2=\frac{(d+2)(d-1)}{2}$ because the $i_1, i_2$ indices transform as a spin-2 representation of $\SO(d)$. The conservation law means that $P_{i_1}\alpha^{\mu\nu}_{i_1i_2}$ is pure gauge and hence should be excluded from the physical spectrum of quasinormal modes. On the other hand, the primary $\gamma$-modes $\gamma^{\mu\nu}_{i_1j_1,i_2j_2}$ have quasinormal frequency $\omega_{QN}=-2 i$ and degeneracy $D^d_{22}=\frac{1}{12}(d+2)(d+1)d(d-3)$ because the indices $[i_1,j_1], [i_2,j_2]$ transform as Weyl tensor under $\SO(d)$. This also explains the Bianchi identity $P_{[k}  \gamma^{\mu\nu}_{i_1j_1],i_2j_2}=0$.

With the higher spin quasinormal modes known, we define a quasinormal character $\chi^{\text{QN}}_s(q)$, cf. (\ref{quacha}) that encodes precisely the information of quasinormal spectrum. We show that $\chi^{\text{QN}}_s(q)$ is equal to the Harish-Chandra group character $\chi^{\text{HC}}_s(q)$ of the unitary massless spin-$s$ $\SO(1, d+1)$ representation. In other words, the pure group theoretical object $\chi^{\text{HC}}_s(q)$ knows everything about the physical quasinormal spectrum.

Our algebraic approach to quasinormal modes has some potential generalizations and applications which will be left to investigate in the future:
\begin{itemize}
\item Construct quasinormal modes of fields carrying other unitary representations, for example partially massless fields or discrete series fields. The generalization to partially massless fields should be more or less straightforward. In particular, the construction of the primary $\alpha$-modes would be the same except the conservation law being replaced by a multiply-conservation equation \cite{Brust:2016zns}:
\begin{align}
P_{i_1}\cdots P_{i_{s-t}}\alpha^{\mu_1\cdots\mu_s}_{i_1\cdots i_s}=\text{pure gauge}
\end{align}
where $t$ is the depth. For the primary $\gamma$-modes,  the higher spin Weyl tensor used in eq. (\ref{gammas}) is expected to be replaced by its partially massless counterpart which carries $\mY_{s, t+1}$ representation of $\SO(d)$ \cite{Hinterbichler:2016fgl}. The discrete series case should be different because it is labelled by a maximal height Young diagram. As a result, neither of the primary quasinormal modes can be a curvature like object. This is also confirmed from the character side \cite{Basile:2016aen}.

\item Generalize the ``quasinormal quantization''  \cite{Ng:2012xp,Jafferis:2013qia} to massless higher spin gauge  fields in any higher dimension. It's well known that quasinormal modes are nonnormalizable with respect to the standard Klein-Gordon inner product. However, it is noticed in \cite{Ng:2012xp,Jafferis:2013qia} that, at least for light scalar fields in $\text{dS}_4$, there is the so-called ``R-norm'' such that the quasinormal modes become normalizable and $\SO(1,4)$ is effectively Wick rotated  to $\SO(2,3)$. Granting the existence of ``R-norm'' in higher dimensions for massless higher spin fields that maps $\SO(1, d+1)$ to $\SO(2, d)$, then the $\gamma$-tower of quasinormal modes carries the $[\Delta=2, \mY_{ss}]$ representation of $\SO(2, d)$, that is below the unitarity bound for sufficiently large $s$.  This simple argument  seems to question the naive generalization of  ``R-norm''.

\item In this paper, we've focused on  Harish-Chandra character $\chi_R(q)=\tr_R\, q^D$ with only the  scaling operator $D$ turned on, where $R$ denotes  some unitary irreducible representation. In general, we can also include $\SO(d)$ generators in the definition of characters:
\begin{align}\label{full}
\chi^{\text{HC}}_R(q, x)=\tr_R \,\left(q^D x_1^{J_1}\cdots, x_r ^{J_r}\right), \,\,\,\,\, r=\floor*{\frac{d}{2}}
\end{align}
where $J_i=L_{2i-1,2i}$ span the Cartan algebra of $\SO(d)$ and $x_i$ are auxiliary variables. For example, for spin-$s$ principal series or complementary series, the full character (\ref{full}) reads \cite{Basile:2016aen,Hirai:Char}
\begin{align}
\chi^{\text{HC}}_{[\Delta,s]}(q,x)=(q^\Delta+q^{\bar\Delta}) \chi^{\SO(d)}_{\mY_{s}}(x)P_d (q, x)
\end{align}
where $\chi^{\SO(d)}_{\mY}(x)\equiv \tr_{\mY} \,x_1^{J_1}\cdots x_r^{J_r}$ denotes the $\SO(d)$ character of spin-$\mY$ representation and
\begin{align}
 P_d (q, x)=\frac{1}{\prod_{i=1}^r(1-x_i q)(1-x^{-1}_i q)}\times\begin{cases} 1, \,\,\,\,\, & d=2r\\ \frac{1}{1-q}, \,\,\,\,\, & d=2r+1 \end{cases}
\end{align}
For massless spin-$s$ representation, the full character originally computed in  \cite{Hirai:Char} is:
\small
\begin{align}
d=2r+1:\,\,\,\,\chi^{\text{HC}}_s (q,x)&= 2\,\left(\chi^{\SO(d)}_{\mY_s} (x) \,q^{s+d-2}-\chi^{\SO(d)}_{\mY_{s-1}}(x)\, q^{s+d-1}\right)P_d(q,x)\nonumber\\
&+\sum_{n=2}^{r} (-)^{n} (q^n-q^{d-n}) \chi^{\SO(d)}_{\mY_{(ss,1^{n-2})}}(x) \, P_d(q,x)
\end{align}
\normalsize
and
\small
\begin{align}
d=2r:\,\,\,\,\chi^{\text{HC}}_s (q,x)&=\sum_{n=2}^{r-1} (-)^{n} (q^n+q^{d-n}) \chi^{\SO(d)}_{\mY_{(ss,1^{n-2})}}(x)P_d(q,x)\nonumber\\
&+(-)^r \, q^r\, \left( \chi^{\SO(d)}_{\mY_{(ss1,\cdots,+1)}}(x)+\chi^{\SO(d)}_{\mY_{(ss1,\cdots,-1)}}(x)\right)P_d(q,x)
\end{align}
\normalsize
We conjecture that the full Harish-Chandra character encodes spin content of quasinormal modes. More precisely, expand $\chi_s^{\text{HC}}(q, x)$ in terms of $q$ and $x_i$
\begin{align}
\chi_s^{\text{HC}}(q, x)=\sum_{\omega, \,\vec j} \, d_{\omega, \vec j} \, q^{i\omega}\, x_1^{j_1}\cdots x_r^{j_r}, \,\,\,\,\, \vec j=(j_1,\cdots, j_r)
\end{align} 
then $d_{\omega, \vec j}$ is conjectured to be the degeneracy of quasinormal modes with frequency $\omega$ and spin content $\vec j$.
\end{itemize}

\section*{Acknowledgments} 
I'm grateful to Frederik Denef, Austin Joyce and Albert Law for numerous stimulating discussions, at different stages of this research project.
I also thank Dionysios Anninos and Frederik Denef for reading the paper and providing precious comments.
ZS was supported in part by the U.S. Department of Energy grant de-sc0011941.

\appendix

\section{From ambient space to intrinsic coordinate:  Maxwell field }\label{Max2}
In this appendix, we show the agreement between our algebraically constructed primary quasnormal modes and their intrinsic coordinate counterparts in literature for free Maxwell fields. According to \cite{LopezOrtega:2006my}, the quasinormal modes of Maxwell theory can be divided into the following two types:
\small
\begin{align} 
\text{\rom{1}}: A_t^{(\rom{1})}=0, \,\,\,\,\, A^{(\rom{1})}_r= R^{(\rom{1})}(r) Y^{\ell \sigma} e^{-i\omega t},\,\,\,\,\, A_a^{(\rom{1})}= \frac{r^{3-d}(1-r^2)}{\ell(\ell+d-2)}\partial_r(r^{d-1}\,R^{(\rom{1})}(r)) \partial_{\vartheta^a}Y^{\ell \sigma}e^{-i\omega t}
\end{align}
\normalsize
where $\vartheta^a$ are the spherical coordinates on $S^{d-1}$ and $Y^{\ell\sigma}$ are scalar spherical harmonics with $\sigma$ being a collective symbol for the magnetic quantum numbers, 
\begin{align} 
\text{\rom{2}}: A_t^{(\rom{2})}=A^{(\rom{2})}_r= 0, \,\,\,\,\, A_a^{(\rom{2})}= R^{(\rom{2})}(r) Y_a^{\ell \sigma}e^{-i\omega t}
\end{align}
where $Y^{\ell \sigma}_a$ are divergence-free vector spherical harmonics on $S^{d-1}$. In type \rom{1} solutions, the radial function $R^{(\rom{1})}(r)$ is given by 
\begin{align}\label{R1}
R^{(\rom{1})}(r)=r^{\ell-1}(1-r^2)^{\frac{i\omega}{2}} F\left(\frac{\ell+i\omega+d-2}{2}, \frac{\ell+i\omega+2}{2}, \frac{d}{2}+\ell,r^2\right)
\end{align}
with the quasinormal frequency $\omega$ valued in
\begin{align}\label{t1}
i\omega^{\rom{1}}_{\ell, n}=\ell+d-2+2n,\,\,\, i\tilde \omega^{\rom{1}}_{\ell,n}=\ell+2+2n
\end{align}
In type \rom{2} solutions, the radial function $R^{(\rom{2})}(r)$ is given by 
\begin{align}\label{R2}
R^{(\rom{2})}(r)=r^{\ell+1}(1-r^2)^{\frac{i\omega}{2}} F\left(\frac{\ell+i\omega+d-1}{2}, \frac{\ell+i\omega+1}{2}, \frac{d}{2}+\ell,r^2\right)
\end{align}
with the quasinormal frequency $\omega$ valued in
\begin{align}\label{t2}
i\omega^{\rom{2}}_{\ell, n}=\ell+d-1+2n,\,\,\, i\tilde \omega^{\rom{2}}_{\ell,n}=\ell+1+2n
\end{align}
In both type \rom{1} and \rom{2}, $\ell\ge 1$ and $n\ge 0$. In the following, we show that the primary quasinormal modes $\alpha^{(1)}_i$ agree with the type \rom{1} solutions of frequency $i\omega^{\rom{1}}_{1,0}$ and $\gamma^{(1)}_{ij}$ agrees with the type \rom{2} solutions of frequency $i\tilde \omega^{\rom{2}}_{1,0}$.

\noindent{}\textbf{Match the primary $\alpha^{(1)}_i$-mode}

\noindent{}Using eq. (\ref{R1}), the type \rom{1} quasinormal modes with $\ell=1$ and $i\omega=i\omega^{\rom{1}}_{1,0}=d-1$ are
\begin{align}\label{A1}
A^{(\rom{1})}_t=0, \,\,\,\,\, A^{(\rom{1})}_r=\frac{e^{-(d-1)t}}{(1-r^2)^{\frac{d-1}{2}}} Y^{1\sigma}(\Omega),\,\,\,\,\, A^{(\rom{1})}_a=\frac{r\, e^{-(d-1)t}}{(1-r^2)^{\frac{d-1}{2}}} \partial_{\vartheta^a}Y^{1\sigma}(\Omega)
\end{align}
On the other hand, the pull-back of $\alpha^{(1)}_i=\frac{X^+ u_i-U^+ x_i}{(X^+)^d} R^{d-2}$ yields
\begin{align}
&\alpha^{(1)}_{i, t}=-\frac{x_i \partial_t X^+}{(X^+)^d}=-\frac{r\,\Omega_i\, e^{-(d-1)t}}{(1-r^2)^{\frac{d-1}{2}}}\nonumber\\
&\alpha^{(1)}_{i, r}=\frac{X^+\partial_r x_i - x_i \partial_r X^+}{(X^+)^d}=\frac{\Omega_i\, e^{-(d-1)t}}{(1-r^2)^{\frac{d+1}{2}}}\nonumber\\
&\alpha^{(1)}_{i, a}=\frac{X^+\partial_{\vartheta^a}x_i}{(X^+)^d}=\frac{r\,\partial_{\vartheta^a}\Omega_i\, e^{-(d-1)t}}{(1-r^2)^{\frac{d-1}{2}}}
\end{align}
Naively, $A^{(\rom{1})}_\mu$ and $\alpha_{i,\mu}$ look different. This is because the former is solved in a modified Feynman gauge \cite{Crispino:2000jx} while the latter follows  from boundary-to-bulk propagator in de Donder gauge, which for spin-1 field is simply the Lorenz gauge. To compare the two results, we perform a gauge transformation $\alpha^{(1)}_{i,\mu}\to \tilde\alpha^{(1)}_{i,\mu}=\alpha_{i,\mu}+\partial_\mu\xi_i$ to set the $t$-component zero. The simplest choice of the gauge parameter is $\xi_i=\frac{1}{d-1}\alpha_{i, t}$. Due to this gauge choice, the new $\alpha^{(1)}_i$ modes become
\begin{align}\label{alphaxi}
\tilde\alpha^{(1)}_{i, t}=0,\,\,\,\,\,\tilde\alpha^{(1)}_{i, r}=\frac{d-2}{d-1}\frac{\Omega_i\, e^{-(d-1)t}}{(1-r^2)^{\frac{d-1}{2}}} ,\,\,\,\,\,\tilde\alpha^{(1)}_{i,a}=\frac{d-2}{d-1}\frac{r\, \partial_{\vartheta^a}\Omega_i\,e^{-(d-1)t}}{(1-r^2)^{\frac{d-1}{2}}} 
\end{align}  
Since $\{\Omega_i\}_i$ and $\{Y^{1\sigma}\}_\sigma$ are just different basis for the same vector space of spherical harmonics of eigenvalue $-(d-1)$ with respect to $\nabla^2_{S^{d-1}}$, eq.(\ref{A1}) and eq. (\ref{alphaxi}) actually represent the same set of quasinormal modes.

\noindent{}\textbf{Match the primary $\gamma^{(1)}_{ij}$-mode}

\noindent{}Using eq. (\ref{R2}), the type \rom{2} quasinormal modes with $\ell=1$ and $i\omega=i\tilde \omega^{\rom{2}}_{1,0}=2$ are
\begin{align}\label{A2}
A^{(\rom{2})}_t=A^{(\rom{2})}_r=0,\,\,\,\,\, A^{(\rom{2})}_a=\frac{r^2\, e^{-2 t}}{1-r^2}Y_a^{1\sigma}(\Omega)
\end{align}
On the other hand, the pull-back of $\gamma^{(1)}_{ij}=\frac{x_i u_j-x_j u_i}{(X^+)^2}$ yields
\begin{align}\label{gamma2}
\gamma^{(1)}_{ij, t}=\gamma^{(1)}_{ij,r}=0, \,\,\,\,\, \gamma^{(1)}_{ij,a}=\frac{r^2\, e^{-2 t}}{1-r^2}\left(\Omega_i\partial_{\vartheta^a}\Omega_j-\Omega_j\partial_{\vartheta^a}\Omega_i\right)
\end{align}
One can check directly that $\Sigma_{ij, a}\equiv \Omega_i\partial_{\vartheta^a}\Omega_j-\Omega_j\partial_{\vartheta^a}\Omega_i$ are indeed divergence-free vector harmonics of $\ell=1$. For example, let's consider the $d=3$ case where the vector harmonics are given by $Y_a^{\ell m}=\frac{1}{\sqrt{\ell(\ell+1)}}\epsilon_{ab}\nabla^b Y^{\ell m}$ \cite{Higuchi:1986wu}:
\begin{align}
Y_a^{1,0}=\frac{1}{2}\sqrt{\frac{3}{2\pi}}(0,\sin^2\theta), \,\,\,\,\, Y_a^{1,\pm 1}=\frac{1}{4}\sqrt{\frac{3}{\pi}}e^{\pm i \varphi}(-i, \pm \sin\theta\cos\theta)
\end{align} 
where $\vartheta^a=(\theta, \varphi)$ are the usual spherical coordinates on $S^2$. Meanwhile, by working out $\Sigma_{ij,a}$ explicitly, we obtain
\begin{align}
\Sigma_{12,a}=(0, \sin^2\theta),\,\,\,\,\, \Sigma_{23,a}\pm i \Sigma_{31,a}=\mp e^{\pm i\varphi}(-i,\pm \sin\theta\cos\theta)
\end{align}
Therefore, $\gamma^{(1)}_{ij,\mu}$ in (\ref{gamma2}) and $A^{(\rom{2})}_\mu$ in (\ref{A2}) represent the same quasinormal modes.

\section{Match quasinormal spectrums}\label{match}
In the section \ref{char}, we defined a quasinormal character $\chi^{\text{QN}}$ for a given quasinormal spectrum $\{\omega, d_\omega\}$, cf. (\ref{quacha}). By definition, the correspondence between quasinormal characters and quasinormal spectrums is one-to-one .
In this appendix, by using quasinormal characters, we show that our algebraic construction yields the same quasinormal spectrum as \cite{LopezOrtega:2006my} for Maxwell fields and linearized gravity.  On the algebraic side, the quasinormal character of a massless spin-$s$ field is shown to be given by eq. (\ref{flipping}). In particular, for $s=1$ and $s=2$, the quasinormal characters read
\begin{align}
\chi^{\text{QN}}_1(q)=\frac{dq^{d-1}-q^d}{(1-q)^d}+\frac{dq-1}{(1-q)^d}+1
\end{align}
and
\begin{align}
\chi^{\text{QN}}_2(q)=\frac{D^d_2\,q^d-D^d_1\,q^{d+1}}{(1-q)^d}+\frac{D^d_2 -D^d_1\,q^{-1}}{(1-q)^d}+d(q+q^{-1})+\frac{d^2-d+2}{2}
\end{align}
where $D^d_1=d, \,\ D^d_2=\frac{1}{2}(d+2)(d-1)$.

\noindent{}\textbf{Maxwell fields}

\noindent{}In the last appendix, we've summarized the quasinormal modes of  Maxwell fields computed in \cite{LopezOrtega:2006my}. Here let's briefly recap the information  about quasinormal frequencies:
\begin{align}\label{spec1}
&\text{type \rom{1}}: \,\,\,\,\, i\omega^{\rom{1}}_{\ell, n}=\ell+d-2+2n,\,\,\, i\tilde \omega^{\rom{1}}_{\ell,n}=\ell+2+2n\nonumber\\
&\text{type \rom{2}}: \,\,\,\,\,i\omega^{\rom{2}}_{\ell, n}=\ell+d-1+2n,\,\,\, i\tilde \omega^{\rom{2}}_{\ell,n}=\ell+1+2n
\end{align}
where $\ell\ge 1$ and $n\ge 0$. For fixed $\ell$ and $n$, each frequency of type \rom{1} quasinormal modes has degeneracy $D^d_{\ell}$ because $\ell$ labels scalar spherical harmonics while
each frequency of type \rom{2} quasinormal modes has degeneracy $D^d_{\ell 1}$ because $\ell$ labels divergence-free vector spherical harmonics. So the quasinormal character associated to the spectrum (\ref{spec1}) is
\begin{align}\label{intrin1}
\chi^{\text{QN}, intrin}_1(q)&\equiv\sum_{\ell=1}^\infty\sum_{n=0}^\infty D^d_\ell\,(q^{i\omega^{\rom{1}}_{\ell, n}}+q^{i\tilde\omega^{\rom{1}}_{\ell, n}})+D^d_{\ell 1}(q^{i\omega^{\rom{2}}_{\ell, n}}+q^{i\tilde \omega^{\rom{2}}_{\ell, n}})\nonumber\\
&=\frac{q^2+q^{d-2}}{1-q^2}\sum_{\ell\ge 1} D^d_\ell \, q^\ell+\frac{q+q^{d-1}}{1-q^2}\sum_{\ell\ge 1}D^d_{\ell 1} \, q^\ell
\end{align}
where the first sum over $\ell$ simply follows from 
\begin{align}\label{00}
\sum_{\ell\ge 0}D^d_{\ell}\,  q^\ell=\frac{1+q}{(1-q)^{d-1}}
\end{align}
The second sum over $\ell$ in (\ref{intrin1}) can be derived using 
\begin{align}\label{magic}
D^d_{\ell s}=D^d_\ell D^{d-2}_s-D^{d}_{s-1} D^{d-2}_{\ell+1}
\end{align}
In particular, when $s=1$, $D^d_{\ell 1}=(d-2)D^d_\ell-D^{d-2}_{\ell+1}$ and hence the second sum reduces to the (\ref{00}) type:
\begin{align}\label{01}
\sum_{\ell\ge 1}\, D^d_{\ell 1} \, q^\ell=(d-2)\frac{1+q}{(1-q)^{d-1}}-\left(\frac{1+q^{-1}}{(1-q)^{d-3}}-q^{-1}\right)
\end{align}
Plugging (\ref{00}) and (\ref{01}) into the quasinormal character (\ref{intrin1}) yields
\begin{align}
\chi^{\text{QN}, intrin}_1(q)=\frac{dq^{d-1}-q^d}{(1-q)^d}+\frac{dq-1}{(1-q)^d}+1=\chi^{\text{QN}}_1(q)
\end{align}
which shows the agreement between our algebraic method and the traditional analytical method on the quasinormal spectrum for Maxwell theory.

\noindent{}\textbf{Linearized gravity}

\noindent{}Quasinormal modes  of linearized gravity are divided into three categories. The three types of fluctuation can be solved simultaneously  by using  the so-called Ishibashi-Kodama equation  \cite{ Natario:2004jd, Kodama:2003kk,LopezOrtega:2006my}. The quasinormal frequencies are:
\begin{align}\label{gravqua}
&\text{Scalar type fluctuation}: i\omega^{S}_{\ell, n}=\ell+d-2+2n,\,\,\, i\tilde \omega^{S}_{\ell,n}=\ell+2+2n\nonumber\\
&\text{Vector type fluctuation}: i\omega^{V}_{\ell, n}=\ell+d-1+2n,\,\,\, i\tilde \omega^{V}_{\ell,n}=\ell+1+2n\nonumber\\
&\text{Tensor type fluctuation}: i\omega^{T}_{\ell, n}=\ell+d+2n,\,\,\, i\tilde \omega^{T}_{\ell,n}=\ell+2n
\end{align}
where $\ell\ge 2$ and $n\ge 0$. In these 3 types of fluctuations, $\ell$ labels  scalar spherical harmonics, divergence-free vector spherical harmonics and divergence-free tensor spherical harmonics on $S^{d-1}$ respectively and hence for fixed $\ell$ and $n$, each frequency has degeneracy $D^d_\ell, D^d_{\ell1}$ and $D^d_{\ell2}$ respectively. Altogether, the quasinormal character associated to the spectrum (\ref{gravqua}) is given by 
\begin{align}\label{intrin2}
\chi^{\text{QN}, intrin}_2(q)&\equiv\sum_{\ell\ge 2, n\ge 0} D^d_\ell\, (q^{i\omega^S_{\ell n}}+q^{i\tilde\omega^S_{\ell n}})\!+\!D^d_{\ell 1}(q^{i\omega^V_{\ell n}}+q^{i\tilde\omega^V_{\ell n}})+D^d_{\ell 2}(q^{i\omega^T_{\ell n}}+q^{i\tilde\omega^T_{\ell n}})\nonumber\\
&=\frac{q^2+q^{d-2}}{1-q^2}\sum_{\ell\ge 2}\, D^d_\ell\, q^\ell+\frac{q+q^{d-1}}{1-q^2}\sum_{\ell\ge 2}\, D^d_{\ell 1}\, q^\ell+\frac{1+q^{d}}{1-q^2}\sum_{\ell\ge 2}\, D^d_{\ell2}\, q^\ell
\end{align}
where the first two series of $\ell$ are essentially computed in the Maxwell field case and the last series follows from eq.(\ref{magic}) with $s=2$:
\begin{align}
\sum_{\ell\ge 2}\, D^d_{\ell2}\, q^\ell=D^d_2\frac{1+q}{(1-q)^{d-1}}-d\left(\frac{1+q^{-1}}{(1-q)^{d-3}}-q^{-1}\right)+\frac{d(d-1)}{2}
\end{align}
Combine the three series of $\ell$  in (\ref{intrin2}) and we obtain 
\begin{align}\label{Q2}
\chi^{\text{QN}, intrin}_2(q)=\frac{D^d_2\, (q^d)-D^d_1\,q^{d+1}}{(1-q)^d}+\frac{D^d_2 -D^d_1\,q^{-1}}{(1-q)^d}+d(q+q^{-1})+\frac{d^2-d+2}{2}
\end{align}
which is exactly  $\chi^{\text{QN}}_2(q)$. This computation confirms the match of quasinormal spectrum for linearized gravity. \footnote{When $d=3$, the tensor type fluctuations doesn't exist because there is no divergence-free tensor harmonics on $S^2$. However, one can still  recover the quasinormal character $\chi^{\text{QN}}_2(q)$ for $d=3$ by counting quasinormal modes in the scalar and vector type fluctuations in this case. }

\section{Details of $\gamma^{(s)}_{i_1 \ell_1,\cdots, i_s \ell_s}$}\label{primariness}
The higher spin  quasinormal mode $\gamma_{i_1 \ell_1,\cdots, i_s \ell_s}^{(s)}$ defined by eq. (\ref{gammas}) is rather schematic. In this appendix, we will write out its explicit form and then show various properties of it. Let's start from recollecting the definitions 
\begin{align}\label{def1}
&\gamma^{(s)}_{i_1 \ell_1,\cdots, i_s \ell_s}(X, U)=\Pi_{ss} P_{i_1}\cdots P_{i_s} (-\log (X^+) \beta^{(s)}_{\ell_1\cdots \ell_s})-\text{trace}\\
&\beta^{(s)}_{\ell_1\cdots \ell_s}(X,U)=\frac{1}{(X^{+})^2}(X^+u_{\ell_1}-U^+ x_{\ell_1})\cdots (X^+u_{\ell_s}-U^+ x_{\ell_s})-\text{trace}\\
& P_i=-X^-\partial_{x^i}-2\,x_i\,\partial_{X^+}-U^-\partial_{u^i}-2\,u_i\,\partial_{U^+}
\end{align}
where the projection operator $\Pi_{ss}$ antisymmetrizes $[i_1, \ell_1],\cdots, [i_s, \ell_s]$. Notice that $X^-\partial_{x^i}$ and $U^-\partial_{u^i}$ would introduce terms proportional $\delta_{i_j i_k}$ and $\delta_{i_j\ell_k}$. The former is killed by $\Pi_{ss}$ and the latter as a pure trace term also drops out in $\gamma^{(s)}_{i_1 \ell_1,\cdots, i_s \ell_s}$. Therefore only $2x_i\partial_{X^+}$ and $2u_i\partial_{U^+}$ can have nonvanishing contributions to $\gamma^{(s)}_{i_1 \ell_1,\cdots, i_s \ell_s}$ and $\gamma^{(s)}_{i_1 \ell_1,\cdots, i_s \ell_s}$ is independent of $X^-$ and $U^-$. As a result, $\gamma^{(s)}_{i_1 \ell_1,\cdots, i_s \ell_s}$ has to be of the following form
\begin{align}
&\gamma^{(s)}_{i_1 \ell_1,\cdots, i_s \ell_s}(X, U)=\chi^{(s)}_{i_1 \ell_1,\cdots, i_s \ell_s}(x, u) f(X^+, U^+)\\
&\chi^{(s)}_{i_1 \ell_1,\cdots, i_s \ell_s}(x, u)\equiv (x_{i_1}u_{\ell_1}-u_{i_1}x_{\ell_1})\cdots (x_{i_s}u_{\ell_s}-u_{i_s}x_{\ell_s})-\text{trace}
\end{align}
where $f(X^+, U^+)$ is an unknown function to be fixed. With $x_i, u_\ell$ being $\SO(d)$ vectors, the tensor $\chi^{(s)}_{i_1 \ell_1,\cdots, i_s \ell_s}(x, u)$ carries the $\mY_{ss}$ representation of $\SO(d)$. In particular, it satisfies the following set of equations which can be thought as the $\SO(d)$ analogue of uplift conditions and Pauli-Fierz conditions:
\begin{align}\label{chicon}
&\partial_x^2\,\chi^{(s)}(x, u)=\partial_u^2\,\chi^{(s)}(x, u)=\partial_x\cdot\partial_u\,\chi^{(s)}(x, u)=0\nonumber\\
x\cdot\partial_u\,\chi^{(s)}(x, u)&=u\cdot\partial_x\,\chi^{(s)}(x, u)=(x\cdot\partial_x-s)\,\chi^{(s)}(x, u)=(u\cdot\partial_u-s)\,\chi^{(s)}(x, u)=0
\end{align}
where the subscripts of $\chi^{(s)}_{i_1\ell_1,\cdots,i_s\ell_s}$ are suppressed.

Using the definition (\ref{def1}), it's easy to check that $\gamma^{(s)}_{i_1 \ell_1,\cdots, i_s \ell_s}$ satisfies the same conditions (\ref{tan})---(\ref{trace}) as $\beta^{(s)}_{\ell_1\cdots \ell_s}$. In particular, the homogeneity condition and the tangentiality condition yields
\begin{align}
X^+\partial_{U^+} f(X^+, U^+)=0, \,\,\,\,\, X^+\partial_{X^+}f(X^+, U^+)=-2 f(X^+, U^+)
\end{align}
which have solution $f(X^+, U^+)=\frac{1}{(X^+)^2}$. Therefore, up to an unimportant normalization factor $c_s$, the explicit form of $\gamma^{(s)}_{i_1 \ell_1,\cdots i_s \ell_s}$ is
\begin{align}\label{gammaexplicit}
\boxed{\gamma^{(s)}_{i_1 \ell_1,\cdots, i_s \ell_s}=c_s\frac{\chi^{(s)}_{i_1 \ell_1,\cdots, i_s \ell_s}(x, u)}{(X^+)^2}=c_s\frac{(x_{i_1}u_{\ell_1}-u_{i_1}x_{\ell_1})\cdots (x_{i_s}u_{\ell_s}-u_{i_s}x_{\ell_s})-\text{trace}}{(X^+)^2}}
\end{align} 
For example, for a Maxwell filed, (\ref{gammaexplicit}) is consistent with (\ref{gamma1}). More generally, all the $\gamma$-primaries that are generated by $\SO(d)$ action on (\ref{gammaexplicit})	can be collectively expressed as 	
\begin{align}\label{gammaindexfree}
\gamma^{(s)}(X, U)=\frac{T_{ss}(x, u)}{(X^+)^2}
\end{align}
where $T_{ss}(x, u)$ is a polynomial in $x^i, u^i$ carrying the $\mY_{ss}$ representation of $\SO(d)$ \footnote{Taking eq. (\ref{gammaindexfree}) as an ansatz of quasinormal modes (which satisfy the in-going boundary condition automatically as we will see below), then it's easy to show that uplift conditions and  Pauli-Fierz conditions are equivalent to $T_{ss}(x, u)$ carrying the $\mY_{ss}$ representation of $\SO(d)$.}, i.e. it satisfies
\begin{align}\label{DefineTSS}
T_{ss}(a x, b u)=(a\,b)^s T_{ss}(x,u), \,\,\,\,\, u\cdot \partial_x T_{ss}(x, u)=\partial_u^2 T_{ss}(x, u)=0
\end{align}
 All the linearly independent choices of $T_{ss}$ correspond to the degeneracy of $\gamma^{(s)}$.

Near horizon $\gamma^{(s)}_{i_1 \ell_1,\cdots, i_s \ell_s}$ becomes singular because $X^+\to 0$. This singular behavior also shows the  following in-going boundary condition: 
\begin{align}
\boxed{\gamma^{(s)}_{i_1 \ell_1,\cdots, i_s \ell_s}(\rho\to \infty)\sim\frac{1}{(X^+)^2}\sim e^{-2(T-\rho)}}
\end{align}
where we can directly read off the quasinormal frequency $i\omega=2$. 

The next task is to show that $\gamma^{(s)}_{i_1 \ell_1,\cdots, i_s \ell_s}$ is primary up to gauge transformation by using its explicit form (\ref{gammaexplicit}) (we can also use (\ref{gammaindexfree}) with $T_{ss}(x, u)$ subject to (\ref{DefineTSS})). Acting the special conformal transformation $K_m$ on $\gamma^{(s)}_{i_1 \ell_1,\cdots, i_s \ell_s}$, we get (dropping the normalization constant $c_s$)
\begin{align}
K_m\,\gamma^{(s)}_{i_1 \ell_1,\cdots, i_s \ell_s}&=\frac{1}{X^+}\partial_{x^m}\chi^{(s)}_{i_1 \ell_1,\cdots, i_s \ell_s}+\frac{U^+}{(X^+)^2}\partial_{u^m}\chi^{(s)}_{i_1 \ell_1,\cdots, i_s \ell_s}\nonumber\\
&=U\cdot\partial_X\left(-\frac{\partial_{u^m}\chi^{(s)}_{i_1 \ell_1,\cdots, i_s \ell_s}}{X^+}\right)+\frac{1}{X^+}\left(u\cdot\partial_x \partial_{u^m}+\partial_{x^m}\right) \chi^{(s)}_{i_1 \ell_1,\cdots, i_s \ell_s}
\end{align}
where we've replaced $U\cdot\partial_X$ by $u\cdot\partial_x$ in the second term of the second line because $\chi^{(s)}_{i_1 \ell_1,\cdots i_s \ell_s}$ only depends on $x^i, u^i$. In addition, noticing that $u\cdot\partial_x$ kills $\chi^{(s)}_{i_1 \ell_1,\cdots i_s \ell_s}$, the product of operators $u\cdot\partial_x\partial_{u^m}$ can be replaced by the corresponding commutator $[u\cdot\partial_x, \partial_{u^m}]=-\partial_{x^m}$ which cancels the other derivative with respect to $x^m$. Altogether, $\gamma^{(s)}_{i_1 \ell_1,\cdots i_s \ell_s}$  is a primary quasinormal mode in the sense that $K_m \gamma^{(s)}_{i_1 \ell_1,\cdots i_s \ell_s}$ can be removed by a gauge transformation
\begin{align}
\boxed{K_m\,\gamma^{(s)}_{i_1 \ell_1,\cdots, i_s \ell_s}(X, U)=U\cdot\partial_X\left(-c_s\frac{\partial_{u^m}\chi^{(s)}_{i_1 \ell_1,\cdots, i_s \ell_s}(x,u)}{X^+}\right)}
\end{align}
where we've restored the normalization constant $c_s$.

\def\refskip{

\newpage{\pagestyle{empty}\cleardoublepage}


\vskip2pt}

\begin{thebibliography}{100}


\bibitem{TheLIGOScientific:2016src} 
  B.~P.~Abbott {\it et al.} [LIGO Scientific and Virgo Collaborations],
  ``Tests of general relativity with GW150914,''
  Phys.\ Rev.\ Lett.\  {\bf 116}, no. 22, 221101 (2016)
  Erratum: [Phys.\ Rev.\ Lett.\  {\bf 121}, no. 12, 129902 (2018)]
  doi:10.1103/PhysRevLett.116.221101, 10.1103/PhysRevLett.121.129902
  [arXiv:1602.03841 [gr-qc]].
  


\bibitem{Natario:2004jd} 
  J.~Natario and R.~Schiappa,
  ``On the classification of asymptotic quasinormal frequencies for d-dimensional black holes and quantum gravity,''
  Adv.\ Theor.\ Math.\ Phys.\  {\bf 8}, no. 6, 1001 (2004)
  doi:10.4310/ATMP.2004.v8.n6.a4
  [hep-th/0411267].

\bibitem{Berti:2009kk} 
  E.~Berti, V.~Cardoso and A.~O.~Starinets,
  ``Quasinormal modes of black holes and black branes,''
  Class.\ Quant.\ Grav.\  {\bf 26}, 163001 (2009)
  doi:10.1088/0264-9381/26/16/163001
  [arXiv:0905.2975 [gr-qc]].


\bibitem{Kokkotas:1999bd} 
  K.~D.~Kokkotas and B.~G.~Schmidt,
  ``Quasinormal modes of stars and black holes,''
  Living Rev.\ Rel.\  {\bf 2}, 2 (1999)
  doi:10.12942/lrr-1999-2
  [gr-qc/9909058].

\bibitem{Konoplya:2011qq} 
  R.~A.~Konoplya and A.~Zhidenko,
  ``Quasinormal modes of black holes: From astrophysics to string theory,''
  Rev.\ Mod.\ Phys.\  {\bf 83}, 793 (2011)
  doi:10.1103/RevModPhys.83.793
  [arXiv:1102.4014 [gr-qc]].
  
\bibitem{Brady:1999wd}
  P.~R.~Brady, C.~M.~Chambers, W.~G.~Laarakkers and E.~Poisson,
  ``Radiative falloff in Schwarzschild-de Sitter space-time,''
  Phys.\ Rev.\ D {\bf 60} (1999) 064003
  doi:10.1103/PhysRevD.60.064003
  [gr-qc/9902010].

  
\bibitem{LopezOrtega:2006my} 
  A.~Lopez-Ortega,
  ``Quasinormal modes of D-dimensional de Sitter spacetime,''
  Gen.\ Rel.\ Grav.\  {\bf 38}, 1565 (2006)
  doi:10.1007/s10714-006-0335-9
  [gr-qc/0605027].
  

\bibitem{LopezOrtega:2012vi} 
  A.~Lopez-Ortega,
  ``On the quasinormal modes of the de Sitter spacetime,''
  Gen.\ Rel.\ Grav.\  {\bf 44}, 2387 (2012)
  doi:10.1007/s10714-012-1398-4
  [arXiv:1207.6791 [gr-qc]].

 

 
\bibitem{Ng:2012xp} 
  G.~S.~Ng and A.~Strominger,
  ``State/Operator Correspondence in Higher-Spin dS/CFT,''
  Class.\ Quant.\ Grav.\  {\bf 30}, 104002 (2013)
  doi:10.1088/0264-9381/30/10/104002
  [arXiv:1204.1057 [hep-th]].


\bibitem{Jafferis:2013qia} 
  D.~L.~Jafferis, A.~Lupsasca, V.~Lysov, G.~S.~Ng and A.~Strominger,
  ``Quasinormal quantization in de Sitter spacetime,''
  JHEP {\bf 1501}, 004 (2015)
  doi:10.1007/JHEP01(2015)004
  [arXiv:1305.5523 [hep-th]].
  
  
 
\bibitem{Tanhayi:2014kba} 
  M.~R.~Tanhayi,
  ``Quasinormal modes in de Sitter space: Plane wave method,''
  Phys.\ Rev.\ D {\bf 90}, no. 6, 064010 (2014)
  doi:10.1103/PhysRevD.90.064010
  [arXiv:1402.2893 [gr-qc]].

  
\bibitem{Basile:2016aen} 
  T.~Basile, X.~Bekaert and N.~Boulanger,
  ``Mixed-symmetry fields in de Sitter space: a group theoretical glance,''
  JHEP {\bf 1705}, 081 (2017)
  doi:10.1007/JHEP05(2017)081
  [arXiv:1612.08166 [hep-th]].
  

\bibitem{Dobrev:1977qv} 
  V.~K.~Dobrev, G.~Mack, V.~B.~Petkova, S.~G.~Petrova and I.~T.~Todorov,
  ``Harmonic Analysis on the n-Dimensional Lorentz Group and Its Application to Conformal Quantum Field Theory,''
  Lect.\ Notes Phys.\  {\bf 63}, 1 (1977).
  doi:10.1007/BFb0009678
  
\bibitem{Hirai:Char} 
T.~Hirai, 
``The characters of irreducible representations of the Lorentz group of n-th order,''
Proc. \ Japan Acad.\ {\bf 41} (1965), no. 7, 526--531
doi:10.3792/pja/1195522333.

  
\bibitem{Hirai:Irrep} 
T.~Hirai, 
`` On irreducible representations of the Lorentz group of n-th order,''
Proc.\ Japan Acad.\ {\bf 38} (1962), no. 6, 258--262
doi:10.3792/pja/1195523378.
  



\bibitem{Anninos:2020hfj}
D.~Anninos, F.~Denef, Y.~T.~A.~Law and Z.~Sun,
``Quantum de Sitter horizon entropy from quasicanonical bulk, edge, sphere and topological string partition functions,''
[arXiv:2009.12464 [hep-th]].


    
\bibitem{Cotton:original} 
  E.~Cotton,
  ``Sur les variet\'es  \`a trois dimensions,''
 Ann. Fac. d. Sc. Toulouse (11) 1, 385 (1899).

\bibitem{Cotton:1} 
  S.J.~Aldersley,
  ``Comments on certain divergence-free tensor densities in a 3-space,''
Journal of Mathematical Physics. 20 (9): 1905-1907.


\bibitem{TD} 
T.~Damour and  S.~Deser, 
`` Geometry of spin 3 gauge theories, ''
 Annales de l'I.H.P. Physique th\'eorique, Volume 47 (1987) no. 3, pp. 277-307.

\bibitem{Henneaux:2015cda} 
  M.~Henneaux, S.~H\"ortner and A.~Leonard,
  ``Higher Spin Conformal Geometry in Three Dimensions and Prepotentials for Higher Spin Gauge Fields,''
  JHEP {\bf 1601}, 073 (2016)
  doi:10.1007/JHEP01(2016)073
  [arXiv:1511.07389 [hep-th]].


\bibitem{Sleight:2017krf} 
  C.~Sleight,
  ``Metric-like Methods in Higher Spin Holography,''
  PoS Modave {\bf 2016}, 003 (2017)
  doi:10.22323/1.296.0003
  [arXiv:1701.08360 [hep-th]].

  

\bibitem{Sleight:2017cax} 
  C.~Sleight and M.~Taronna,
  ``Feynman rules for higher-spin gauge fields on AdS$_{d+1}$,''
  JHEP {\bf 1801}, 060 (2018)
  doi:10.1007/JHEP01(2018)060
  [arXiv:1708.08668 [hep-th]].
  

\bibitem{Mikhailov:2002bp}
A.~Mikhailov,
``Notes on higher spin symmetries,''
[arXiv:hep-th/0201019 [hep-th]].



\bibitem{Costa:2014kfa}
M.~S.~Costa, V.~Gon\c{c}alves and J.~Penedones,
``Spinning AdS Propagators,''
JHEP \textbf{09}, 064 (2014)
doi:10.1007/JHEP09(2014)064
[arXiv:1404.5625 [hep-th]].


\bibitem{Dobrev:1975ru} 
  V.~K.~Dobrev, V.~B.~Petkova, S.~G.~Petrova and I.~T.~Todorov,
  ``Dynamical Derivation of Vacuum Operator Product Expansion in Euclidean Conformal Quantum Field Theory,''
  Phys.\ Rev.\ D {\bf 13}, 887 (1976).
  doi:10.1103/PhysRevD.13.887

\bibitem{Bekaert:2002dt} 
  X.~Bekaert and N.~Boulanger,
  ``Tensor gauge fields in arbitrary representations of GL(D,R): Duality and Poincare lemma,''
  Commun.\ Math.\ Phys.\  {\bf 245}, 27 (2004)
  doi:10.1007/s00220-003-0995-1
  [hep-th/0208058].


\bibitem{Didenko:2014dwa}
  V.~E.~Didenko and E.~D.~Skvortsov,
  ``Elements of Vasiliev theory,''
  arXiv:1401.2975 [hep-th].
  
  
\bibitem{Bousso:2001mw}
R.~Bousso, A.~Maloney and A.~Strominger,
``Conformal vacua and entropy in de Sitter space,''
Phys. Rev. D \textbf{65}, 104039 (2002)
doi:10.1103/PhysRevD.65.104039
[arXiv:hep-th/0112218 [hep-th]].

  
\bibitem{Anninos:2017eib}
  D.~Anninos, F.~Denef, R.~Monten and Z.~Sun,
  ``Higher Spin de Sitter Hilbert Space,''
  arXiv:1711.10037 [hep-th].
  
  
  
  

\bibitem{Dolan:2005wy}
F.~A.~Dolan,
``Character formulae and partition functions in higher dimensional conformal field theory,''
J. Math. Phys. \textbf{47}, 062303 (2006)
doi:10.1063/1.2196241
[arXiv:hep-th/0508031 [hep-th]].

\bibitem{Gibbons:2006ij}
G.~W.~Gibbons, M.~J.~Perry and C.~N.~Pope,
``Partition functions, the Bekenstein bound and temperature inversion in anti-de Sitter space and its conformal boundary,''
Phys. Rev. D \textbf{74}, 084009 (2006)
doi:10.1103/PhysRevD.74.084009
[arXiv:hep-th/0606186 [hep-th]].

  
\bibitem{Brust:2016zns} 
  C.~Brust and K.~Hinterbichler,
  ``Partially Massless Higher-Spin Theory,''
  JHEP {\bf 1702}, 086 (2017)
  doi:10.1007/JHEP02(2017)086
  [arXiv:1610.08510 [hep-th]].
  
  
\bibitem{Hinterbichler:2016fgl} 
  K.~Hinterbichler and A.~Joyce,
  ``Manifest Duality for Partially Massless Higher Spins,''
  JHEP {\bf 1609}, 141 (2016)
  doi:10.1007/JHEP09(2016)141
  [arXiv:1608.04385 [hep-th]].
  
  
  

\bibitem{Crispino:2000jx}
L.~C.~B.~Crispino, A.~Higuchi and G.~E.~A.~Matsas,
``Quantization of the electromagnetic field outside static black holes and its application to low-energy phenomena,''
Phys. Rev. D \textbf{63}, 124008 (2001)
[erratum: Phys. Rev. D \textbf{80}, 029906 (2009)]
doi:10.1103/PhysRevD.80.029906
[arXiv:gr-qc/0011070 [gr-qc]].
  
  
 



  
\bibitem{Higuchi:1986wu}
  A.~Higuchi,
  ``Symmetric Tensor Spherical Harmonics on the $N$ Sphere and Their Application to the De Sitter Group SO($N$,1),''
  J.\ Math.\ Phys.\  {\bf 28} (1987) 1553
   Erratum: [J.\ Math.\ Phys.\  {\bf 43} (2002) 6385].
  doi:10.1063/1.527513
  
\bibitem{Kodama:2003kk} 
  H.~Kodama and A.~Ishibashi,
  ``Master equations for perturbations of generalized static black holes with charge in higher dimensions,''
  Prog.\ Theor.\ Phys.\  {\bf 111}, 29 (2004)
  doi:10.1143/PTP.111.29
  [hep-th/0308128].




\end{thebibliography}
\end{document}